\documentclass[11pt,a4paper,reqno]{amsart}%
\usepackage{amsthm,amsmath,amsfonts,amssymb,appendix,bookmark,delarray,dsfont,color,mathrsfs,soul}
\usepackage[utf8]{inputenc}

\usepackage{bbm} 
\usepackage{hyperref} 

\usepackage{geometry}
\newtheorem{theorem}{Theorem}[section]
\newtheorem{conjecture}[theorem]{Conjecture}

\newtheorem{lemma}[theorem]{Lemma}

\theoremstyle{definition}

\theoremstyle{remark}

\newtheorem{remark}[theorem]{Remark}

\newcommand{\1}{\mathbbm{1}}

\renewcommand{\epsilon}{\varepsilon}

\newcommand{\N}{\mathbb{N}}

\renewcommand{\phi}{\varphi}
\newcommand{\R}{\mathbb{R}}

\newcommand{\cL}{\mathcal{L}}
\newcommand{\cK}{\mathcal{K}}
\newcommand{\cH}{\mathcal{H}}
\newcommand{\cQ}{\mathcal{Q}}

\newcommand{\cN}{\mathcal{N}}

\DeclareMathOperator{\ran}{ran}

\DeclareMathOperator{\Tr}{Tr}
\DeclareMathOperator{\tr}{Tr}

\numberwithin{equation}{section}

\begin{document}
	
	\title[Optimizers for the mass--supercritical Lieb--Thirring Inequalities]{Finite-Rank Optimizers for the mass--supercritical Lieb--Thirring and Hardy--Lieb--Thirring Inequalities}
	
	\author[G.K. Duong]{Giao Ky Duong}
	\address{Department of Mathematics, LMU Munich, Theresienstrasse 39, D-80333 Munich, and Munich Center for Quantum Science and Technology (MCQST), Schellingstr. 4, D-80799 Munich, Germany}
	\email{duong@math.lmu.de}

	\author[T.M.T. Le]{Thi Minh Thao Le}
	\address{Department of Mathematics and Statistics, Masaryk University, Kotlářská 2, 61137 Brno, Czech Republic}
	\email{tmtle@math.muni.cz}
	
	\author[P.T. Nam]{Phan Th\`anh Nam}
	\address{Department of Mathematics, LMU Munich, Theresienstrasse 39, D-80333 Munich, and Munich Center for Quantum Science and Technology (MCQST), Schellingstr. 4, D-80799 Munich, Germany}
	\email{nam@math.lmu.de}
	
	\author[P.T. Nguyen]{Phuoc-Tai Nguyen}
	\address{Department of Mathematics and Statistics, Masaryk University, Kotlářská 2, 61137 Brno, Czech Republic}
	\email{ptnguyen@math.muni.cz}

	\begin{abstract} We establish the existence of finite-rank operators for an interpolation version of the Lieb--Thirring inequality in the mass--supercritical case, thereby extending a result of Hong, Kwon, and Yoon \cite{HKY2019} to the full parameter regime. Our method also applies to the Hardy--Lieb--Thirring inequality, where the existence of optimizers faces additional difficulties due to the singularity of the inverse-square potential.
	\end{abstract}
		\maketitle

	\section{Introduction}
	
	The Lieb--Thirring inequality \cite{LiebThirring75,LiebThirring76} states that  for every self-adjoint operator $0\le \gamma \le 1$ on $L^2(\R^d)$, the kinetic bound
	\begin{equation}
		\Tr(-\Delta \gamma ) \ge K_{\rm LT}(d)  \int_{\R^d} \rho_\gamma(x)^{1 + \frac{2}{d}} \, dx
		\label{standard-LT}
	\end{equation}
	holds with a universal constant  $K_{\rm LT}(d) \in (0,\infty)$ depending only on the dimension $d\ge 1$.  Here the density $\rho_\gamma(x)=\gamma(x,x)$, with $\gamma(x,y)$ being the integral kernel of $\gamma$, can be defined properly via a spectral decomposition if $\Tr(-\Delta \gamma ):= \Tr(\sqrt{-\Delta}\gamma \sqrt{-\Delta})<\infty$ (for instance, if $\gamma=\sum_{n\ge 1} |u_n\rangle \langle u_n|$, then $\rho_\gamma(x)=\sum_{n\ge 1} |u_n(x)|^2$). 
	
	In the case of a rank-one operator $\gamma=|u\rangle\langle u|$, the bound \eqref{standard-LT} implies the \textit{mass--critical}  Gagliardo--Nirenberg interpolation inequality (up to a constant factor)
	\begin{equation}
		\| \nabla u\|_{L^2(\R^d)}^2  \ge K_{\rm GN}(d)  \int_{\R^d} |u(x)|^{2( 1 + \frac{2}{d})} \, dx,\quad \forall u\in H^1(\R^d), \quad \|u\|_{L^2(\R^d)}=1. 
		\label{standard-GN}
	\end{equation}
	From a physical point of view, while the Gagliardo--Nirenberg inequality can be interpreted as a quantitative formulation of the uncertainty principle (similar to the Sobolev inequality), the Lieb--Thirring inequality goes further by also incorporating the exclusion principle, encoded in the assumption $0 \le \gamma \le 1$. In particular, a key feature of \eqref{standard-LT} is that it applies to operators of arbitrary finite or even infinite rank. In  the infinite-rank case, 
	\eqref{standard-LT} is consistent (up to a constant factor) with the semiclassical approximation of the kinetic energy
	\begin{equation*}
		\Tr(-\Delta \gamma ) \approx K_{\rm cl} (d)  \int_{\R^d} \rho_\gamma(x)^{1 + \frac{2}{d}} \, dx,
	\end{equation*}
	which plays a central role in density functional theory. The classical constant
	\begin{equation*}
		K_{\rm cl} (d) = \frac{d}{d+2}  \frac{(2\pi)^2}{|B_1|^{2/d}},
	\end{equation*}
	where $|B_1|$ denotes the volume of the unit ball in $\R^d$, arises from Thomas--Fermi theory, see e.g. \cite{Lieb1981}. 
	We refer to \cite{FLW22} for a textbook of the Lieb--Thirring inequality and to \cite{LS10} for a pedagogical discussion of its applications to the stability of matter in large fermionic quantum systems.
	
	Historically, the original proof of \eqref{standard-LT} by Lieb and Thirring in \cite{LiebThirring75,LiebThirring76} is based on the dual form
	\begin{equation} \label{eq:LT-inequality-dual}
		\Tr(-\Delta - V(x))_{-} \ge -L(d) \int_{\R^d} V_{+}(x)^{1 + \frac{d}{2}}  dx,
	\end{equation}
	where the constant $L(d)$ is related to $K_{\rm LT}(d)$ as 
	\begin{equation} \label{eq:constant-Ld}
		L(d) = \Big(1 + \tfrac{d}{2}\Big)^{-1} \Big[ \Big(1 + \tfrac{2}{d}\Big) K_d \Big]^{-\frac{d}{2}}.
	\end{equation}
	Here, the left-hand side of \eqref{eq:LT-inequality-dual} is the sum of the negative eigenvalues of the Schr\"odinger operator $-\Delta - V(x)$ on $L^2(\R^d)$. The advantage of this dual formulation is that it can be derived by applying the Birman--Schwinger principle to compact operators. Later, direct proofs of the kinetic inequality \eqref{standard-LT} have been found in \cite{EF1991}, \cite{DLL2008}, \cite{Rumin2011,Rumin2010}; we also refer to \cite{Nam-review} for a review.

	Determining the optimal constant $K_{\rm LT}(d)$ in \eqref{standard-LT} is a central question in mathematical physics. Lieb and Thirring conjectured in \cite{LiebThirring76} that \eqref{standard-LT} holds with 
	\begin{equation}\label{conjecture-LT-76} K_{\rm LT}(d) = \left\{ \begin{array}{lll} K_{\rm GN}(d) &\text{ if } d = 1,2, \\[0.2cm]  K_{\rm cl} (d) &\text{ if } d \ge 3, \end{array} \right. 
	\end{equation} 
	or equivalently \eqref{eq:LT-inequality-dual} holds with 
	\begin{equation*} 
    L(d) = \left\{ \begin{array}{lll} L_{\rm GN}(d) &\text{ if } d =1,2, \\[0.2cm] L_{\rm cl}(d) &\text{ if } d \ge 3, \end{array} \right. 
	\end{equation*}
	where $L_{\rm GN}(d)$ and $L_{\rm cl}(d)$ are determined from $K_{\rm GN}(d)$ and $K_{\rm cl}(d)$ via the dual relation \eqref{eq:constant-Ld}. Despite many considerable efforts, with \cite{FHJN-21} providing the best known constant to date, the conjecture remains open in all dimensions. We also refer to \cite{Lev14} for numerical investigations.
	
	Related to the Lieb--Thirring conjecture, it is natural to expect that the trace inequality \eqref{eq:LT-inequality-dual} admits an optimal potential $V$ when $d=1$ or $2$, whereas no such optimizer exists for $d \ge 3$. However, proving such a result remains out of reach with current techniques. A remarkable step in this direction was made recently by Frank, Gontier and Lewin \cite{FGL-25}, who studied  optimizers for finite-rank Lieb--Thirring inequalities. Among other results, they established the existence of optimizers for a variant of \eqref{eq:LT-inequality-dual} involving the sum of the $N$ lowest eigenvalues. In fact, the results in \cite{FGL-25} concern not only the standard sum of eigenvalues but also the Riesz means of eigenvalues (see also \cite{FGL21b} for earlier related work). Nevertheless, in the optimal scenario of \eqref{eq:LT-inequality-dual} (if it exists), it remains unknown whether the optimizer would generate finitely or infinitely many eigenvalues.
	
	Since the Lieb--Thirring inequality corresponds to the mass--critical case, which is very difficult, it is natural to ask analogous questions about the existence of optimizers in the mass--subcritical and mass--supercritical regimes. In this context, it is more convenient to work with the kinetic formulation \eqref{standard-LT}.

	The mass--subcritical case was studied by Gontier, Lewin, and Nazar \cite{GLQ-21}. In this case, it is necessary to introduce an additional mass constraint. To be precise, for every $N\ge 2$, they considered the existence of an optimal projection $\gamma = \sum_{n=1}^N |u_n\rangle \langle u_n|$ (a ground state), where the orthonormal functions ${u_n}$ solves the nonlinear equation 
	\begin{align}\label{eq:LT-interpolation-EL}
		(-\Delta - \rho_\gamma^{q-1}) u_n = -\mu_n u_n, 
	\end{align}
	corresponding to the $N$ lowest eigenvalues $-\mu_1<-\mu_2<...\le -\mu_N\le 0$ of the self-consistent operator $-\Delta - \rho_\gamma^{q-1}$ on $L^2(\R^d)$. They conjectured that in the mass--subcritical regime
	$$
	1< q< 1+\frac{2}{d},
	$$
	a ground state exists for every $N \ge 2$. However, they were only able to prove this conjecture for $q$ sufficiently close to $1$ \cite[Theorem 3]{GLQ-21}, and obtained partial results for the full range $1 < q < 1 + \frac{2}{d}$ \cite[Theorem 4]{GLQ-21} (which in particular implies that there is no global ground state without the mass/rank constraint). 
	
	On the other hand, Hong, Kwon, and Yoon \cite{HKY2019} studied the mass--supercritical (and energy--subcritical) case $1 + \tfrac{2}{d} < q < 1 + \tfrac{2}{(d-2)_+}$. They proved that the corresponding problem always has (at least) one global ground state (without any mass constraint); moreover, if $d \ge 3$ and 
	\begin{align}\label{eq:supercritical-intro}1 + \tfrac{2d+4}{d^2} < q < 1 + \tfrac{2}{d-2},
	\end{align}
	then every ground state has finite trace. The lower bound condition on $q$ arises from an application of the Cwikel--Lieb--Rozenblum (CLR) inequality \cite[Theorem XIII.12]{RS1978} for the number of negative eigenvalues of the operator $-\Delta - \rho_\gamma^{q-1}$, together with the interpolation estimate
	$$\|\rho_\gamma^{q-1}\|_{L^{\frac{d}{2}}(\R^d)} \lesssim \max\{\|\rho_\gamma\|_{L^{1+\frac{2}{d}}(\R^d)}, \|\rho_\gamma\|_{L^{1+\frac{2}{d-2}}(\R^d)}\}.$$ 
	
	In summary, building on the results of \cite{GLQ-21,HKY2019}, it is natural to formulate the following conjecture, which emphasizes the sharp dichotomy between the mass--subcritical and mass--supercritical cases.
	
	\begin{conjecture}[Optimizers of Lieb--Thirring interpolation inequalities]\label{con:1}
		Consider the ground state problem for the nonlinear equation \eqref{eq:LT-interpolation-EL} in dimension $d\ge 2$  under the constraint $0 \le \gamma \le 1$.
		\begin{itemize}
			\item In the mass--subcritical case $q < 1 + 2/d$, for every $N \ge 2$ there exists a ground state solution of rank $N$.
			
			\item In the mass--supercritical case $q > 1 + 2/d$, there exists a finite integer $N_0$ such that no ground state of rank $N$ exists for $N > N_0$. In particular, there exists a global optimizer of finite rank.
		\end{itemize}
	\end{conjecture}
	
	We do not include the one-dimensional case in the above conjecture. This case is special: it was proved in \cite{FGL21b} that for $d=1$ and $q=2$, there is no rank-$N$ optimizer for any $N \ge 2$, and it was conjectured in \cite{GLQ-21} that this nonexistence extends to the full mass--subcritical regime $q \in [2,3)$ in one dimension. Similarly, the behavior in the mass--supercritical regime in one dimension remains unclear. From a technical perspective, all known methods for proving the existence of finite-rank optimizers in the mass--supercritical case rely on CLR-type inequalities, which typically require $d \ge 3$ (or at least $d \ge 2$). 
	
	The aim of the present paper is to address the second part of the above conjecture, namely, to establish the existence of finite-rank optimizers throughout the full mass--supercritical regime. Our approach differs substantially from that of \cite{HKY2019}: we first construct optimizers under the additional finite-trace assumption, and subsequently derive uniform estimates that allow us to remove this constraint at the end. 
	
	Interestingly, our new approach also makes it possible to investigate the Hardy--Lieb--Thirring (HLT) interpolation inequality. In this setting, the ground state equation takes the form (c.f. \eqref{eq:LT-interpolation-EL})
	\begin{align}\label{eq:HLT-interpolation-EL}
		(-\Delta - c_*|x|^{-2}- \rho_\gamma^{q-1}) u_n = -\mu_n u_n,
	\end{align}
	with $d \ge 3$ and $c_*=(d-2)^2/4$. The existence of optimizers for the HLT problem in the mass--supercritical case cannot be deduced directly from the analysis in \cite{HKY2019}. A major difficulty comes from the critical singularity of the Hardy potential, which creates substantial obstacles for both compactness arguments and the study of the associated Euler--Lagrange equation.

	The main results together with the key ideas of their proofs will be presented in the next section. Since our approach is quantitatively robust, we present and prove all results in the framework of the fractional Laplacian.

	\section{Main results}
	
	\subsection{Mass--supercritical Lieb--Thirring inequality}

	For every $s>0$, we denote by $(-\Delta)^s$ the fractional Laplace operator on $L^2(\R^d)$ which is  defined as the multiplication operator $|2\pi k|^{2s}$ in the Fourier space.  In particular, if $s\in (0,1)$, then the fractional Laplacian $(-\Delta)^s$ can be equivalently defined via the following  quadratic form 
	\begin{equation*}\label{defi-semi-norm}
		\left\langle (-\Delta)^s u,u \right\rangle 
		= 
		\frac{a_{d,s}}{2}  \int_{\R^d} \int_{\R^d}  \frac{|u(x) -u(y)|^2}{|x-y|^{d+2s}} d x d y, 
		\,\,{a}_{d,s} 
		=  
		2^{2s} \frac{\Gamma(\frac{d+2s}{2})}{\pi^{\frac{d}{2}}|\Gamma(-s)|},
	\end{equation*}
	with the form domain $\dot{H}^s(\R^d)$. 	For every self-adjoint bounded operator $\gamma\ge 0$ on $L^2(\R^d)$, we denote 
	$$
	\| \gamma\|_{\dot{\mathcal H}^s} = \Tr ( (-\Delta)^s \gamma) := \Tr ((-\Delta)^{s/2} \gamma (-\Delta)^{s/2}) \in [0,\infty]. 
	$$
	
	For every $d\ge 1$ and $s\in (0,1]$, we consider the mass--supercritical (and energy-subcritical) exponent 
	\begin{equation}
		\left\{
		\begin{array}{ll}
			\displaystyle 1 + \frac{2s}{d} \le q < \infty & \text{ if  }  2s \ge d, \\[0.2cm]
			\displaystyle 1+  \frac{2s}{d} \le q \le \frac{d}{d-2s} & \text{  if  }  2s < d.
		\end{array}
		\right.
		\label{q-range-full}
	\end{equation}
	We are interested in the Lieb--Thirring interpolation inequality 
	\begin{equation}
		\|\gamma\|_{\mathrm{op}}^{1 - \theta} \|\gamma\|_{\dot{\mathcal H}^s}^\theta \ge C_{\rm LT} \|\rho_\gamma\|_{L^q(\R^d)}
		\label{LT-inequality}
	\end{equation}
	with 
	\begin{align}\label{theta-range}
		\theta = \frac{d(q-1)}{2sq} \in \left[ \frac{d}{d+2s}, 1 \right],
	\end{align} 
and the sharp constant $C_{\rm LT} = C_{\rm LT}(d,s,q)>0$  independent of the operator $\gamma$. By a standard scaling argument, it suffices to restrict \eqref{LT-inequality} to operators satisfying $0 \leq \gamma \leq 1$. 
	
	The interpolation inequality \eqref{LT-inequality} follows from the corresponding results at the two endpoint cases: the endpoint case $q = 1 + 2s/d$ recovers the classical (fractional) Lieb--Thirring inequality \cite{Lieb1983} (see also \cite{Nam-review} and \cite{Rumin2010}), whereas the endpoint case $q = d/(d-2s)$, if $d>2s$, yields the (fractional) Sobolev inequality, see \cite{Chemin-Xu97} (the latter being a consequence of the Hoffmann--Ostenhof inequality \cite{HOPRA77} $\|\gamma\|_{\dot{\mathcal H}^s} \geq \|\sqrt{\rho_\gamma}\|_{\dot H^s}$). 
	
	In the non-relativistic case $s=1$, the mass--supercritical Lieb--Thirring inequality \eqref{LT-inequality} was investigated by Hong, Kwon, and Yoon in \cite{HKY2019}. They proved that for every $q$ in the non-endpoint regime, there exists an optimizer with finite kinetic energy $\|\gamma\|_{\dot {\cH^1}}<\infty$. Furthermore, they showed that such optimizers must be of finite rank provided $d \ge 3$ and $q > 1 + \frac{2d+4}{d^2}$. 
	
	Our first result provides a construction of finite-rank optimizers for \eqref{LT-inequality} throughout the whole mass--supercritical regime \eqref{q-range-full}, excluding only the endpoint cases. In particular, it addresses the second point of Conjecture \ref{con:1}. 
	
	\begin{theorem}[Finite-rank optimizers for the mass--supercritical Lieb--Thirring inequality]\label{thm:optimizer-LT}
		Let $d \ge 1$, $ 0 < s  \le 1$, $s<d/2$ and $q \in \left(\tfrac{d+2s}{d}, \tfrac{d}{d-2s}\right)$. Then the Lieb--Thirring interpolation inequality \eqref{LT-inequality} has a finite-rank optimizer $0 \le \gamma_\infty \le 1$. This optimizer can be represented in the form $\gamma_\infty = \sum_{n=1}^M |u_n\rangle \langle u_n|$, where $M \in \N$ and $\{u_n\}_{n\geq 1}^M \subset H^s(\R^d)$ is an orthogonal family of functions in $L^2(\R^d)$ satisfying the Euler--Lagrange equation
		$$
			((-\Delta)^s - \rho_{\gamma_\infty}^{q-1}) u_n = -\mu_n u_n, \quad n=1,2,\dots,M,
		$$
		with $-\mu_1 \leq -\mu_2 < \cdots \le - \mu_M \le 0$ denoting the lowest eigenvalues (counting multiplicities) of the operator $(-\Delta)^s - \rho^{q-1}_{\gamma_\infty}$.	Moreover, there exists a finite integer $N_0=N_0(d,s,q)$ such that, for every $N > N_0$, there is no optimizer for \eqref{LT-inequality} among rank-$N$ operators, nor among operators satisfying $\Tr \gamma = N$. 
	\end{theorem}
	
	We hope that our results will stimulate further contributions to the ongoing program on Lieb--Thirring interpolation inequalities, initiated in the mass--subcritical case by \cite{GLQ-21} and further advanced in the mass--supercritical case by \cite{HKY2019}. In particular, Theorem \ref{thm:optimizer-LT} provides a full confirmation of the second part of Conjecture \ref{con:1} in dimensions $d\ge 3$. The first part of Conjecture \ref{con:1}, proposed in  \cite{GLQ-21} concerning the mass--subcritical case, however, remains open.
	
	Our proof of Theorem \ref{thm:optimizer-LT} for $s=1$ is based on the observation that $\|\rho_\gamma^{q-1}\|_{L^{d/2}(\R^d)}$, which comes from the application of the Cwikel--Lieb--Rozenblum (CLR) inequality to estimate the number of negative eigenvalues of the  self-consistent operator $-\Delta - \rho_\gamma^{q-1}$, can be controlled using $\|\rho_\gamma\|_{L^{1+2/(d-2)}(\R^d)}$ and $\|\rho_\gamma\|_{L^1(\R^d)}$ (instead of $\|\rho_\gamma\|_{L^{1+2/d}(\R^d)}$ as used in \cite{HKY2019}). To carry out this strategy, we first restrict the problem to the finite-trace setting $\Tr \gamma \le N$, and then derive uniform estimates in $N$ from the associated Euler--Lagrange equation. These estimates ultimately allow us to remove the finite-trace constraint and thereby obtain a finite-rank optimizer for the original problem. This strategy also applies  to the case $s \in (0,1)$ although the analysis is more intricate.

	\subsection{Mass--supercritical Hardy--Lieb--Thirring inequality}

	In the second part of the paper, we aim to extend  the previous results to  the fractional Hardy  Schr\"odinger  operator.  Let $d \ge 1$ and $s \in \left(0,\frac{d}{2} \right)$, the    Hardy inequality states that
	$$
		\mathcal{L}_s  := (-\Delta)^s  -  \frac{ \mathcal{C}_{d,s}}{|x|^{2s}} \ge 0   
		\quad \text{ on } L^2(\R^d)
	$$
	with  the optimal constant  
	\begin{equation*}
		\mathcal{C}_{d,s}:= 2^{2s} \frac{\Gamma^2\left(\frac{d + 2s}{4}\right)}{\Gamma^2\left(\frac{d - 2s}{4}\right)}. 
	\end{equation*}
	Denote by $Q^s$ the quadratic form domain of $\cL_s$, which is a Hilbert space with the norm
	$$ \| u \|_{Q^s} := \left( \| \sqrt{\cL_s} u \|_{L^2(\R^d)}^2 + \| u \|_{L^2(\R^d)}^2 \right)^{\frac{1}{2}}.
	$$
	For every self-adjoint bounded operator $\gamma\ge 0$, we denote
	$$
	\| \gamma\|_{\dot{\cQ}^s} = \Tr(\mathcal{L}_s \gamma) := \Tr(\sqrt{\mathcal{L}_s} \gamma \sqrt{\mathcal{L}_s }) \in [0,\infty].
	$$
The standard Hardy--Lieb--Thirring inequality asserts that 
	\begin{equation*}
		\| \gamma\|_{\dot{\cQ}^s} \ge K_{\rm HLT}(d,s) \int_{\R^d} \rho_\gamma(x)^{1 + \frac{2s}{d}} dx
	\end{equation*}
	for all $0 \le \gamma \le 1$, with a universal constant $K_{\rm HLT}(d,s) > 0$. This result was established by Frank \cite{F2009} for all $0 < s < d/2$, thereby extending the earlier work of Frank, Lieb, and Seiringer \cite{FLSJAMS08} in the range $0 < s \le 1$, $s < d/2$, as well as the non-relativistic case $s=1$ proved by Ekholm and Frank in \cite{EF2006}.

	\medskip 
	We are interested in the Hardy--Lieb--Thirring interpolation inequality 
	\begin{equation}\label{HLT-inequality-}
		\|\gamma\|_{\mathrm{op}}^{1 - \theta} \| \gamma \|_{\dot{\cQ}^s}^{\theta} \ge C_{\rm HLT} \left\| \rho_\gamma \right\|_{L^q(\R^d)} 
	\end{equation}
	in the mass--supercritical regime 
	\begin{equation} \label{q-range}
		1+\frac{2s}{d} <q < 1+ \frac{2s}{d-2s},
	\end{equation}
	with 	 $\theta \in (0,1) $  defined in \eqref{theta-range} and the sharp constant $C_{\rm HLT}=C_{\rm HLT}(d,s,q)>0$ independent of $\gamma$. By a scaling argument, it suffices to consider \eqref{HLT-inequality-} for $0\le \gamma\le 1$. 	Our second result  reads
	\begin{theorem}[Finite-rank optimizers for the mass--supercritical Hardy--Lieb--Thirring inequality] \label{thm:optfinite}
		Let  $d \ge 1, s \in (0,1], s<d/2$  and $q \in \left( 1+  \frac{2s}{d} , 1+ \frac{2s}{d-2s}  \right)$.  Then the Hardy--Lieb--Thirring interpolation inequality \eqref{HLT-inequality-} has a finite-rank optimizer $0\le \tilde \gamma_\infty \le 1$. This optimizer can be represented in the form $\tilde \gamma_\infty = \sum_{n=1}^{\tilde M} |\tilde u_n\rangle \langle \tilde u_n|$, where $\tilde M \in \N$ and  $\{\tilde u_n\}_{n\geq 1}^{\tilde M}\subset Q^s$ is an orthogonal family of functions in $L^2(\R^d)$ satisfying the Euler--Lagrange equation
		$$
			(\cL_s - \rho_{\tilde \gamma_\infty}^{q-1}) \tilde u_n = -\tilde \mu_n \tilde u_n, \quad n=1,2,\dots,\tilde M,
		$$
		with $-\tilde \mu_1 \leq -\tilde \mu_2 \leq \cdots \le - \tilde \mu_{\tilde M} \le 0$ denoting the lowest eigenvalues (counting multiplicities) of the operator $\cL_s - \rho^{q-1}_{\tilde \gamma_\infty}$. Moreover, there exists a finite integer $ N_1=N_1(d,s,q)$ such that, for every $N > N_1$, there is no optimizer for \eqref{HLT-inequality-} among rank-$N$ operators, nor among operators satisfying $\Tr \gamma = N$. 
	\end{theorem}
	
Note that, unlike the Lieb--Thirring problem \eqref{LT-inequality}, establishing the existence of a global optimizer for the Hardy--Lieb--Thirring inequality in the mass--supercritical case is highly nontrivial, let alone the existence of finite-rank optimizers. In particular, the analysis in \cite{HKY2019} does not seem to extend to the Hardy--Lieb--Thirring problem. Moreover, recently and independently of us, Chen, Guo, and Wu \cite{CGWHLT2025} studied the existence of finite-rank optimizers for the Riesz mean of the first $N$ eigenvalues of the Hardy operator $-\Delta - (d-2)^2/(4|x|^2) - V(x)$. This dual formulation is inspired by earlier works on the Lieb--Thirring problem \cite{FGL21b,FGL-25}, which are related but not identical to the question on the kinetic version that we address in the present paper. 

Our approach to the Hardy--Lieb--Thirring problem is motivated by the analysis of the Lieb--Thirring problem \eqref{LT-inequality}. More precisely, we first establish the existence of an optimizer for the Hardy--Lieb--Thirring problem \eqref{HLT-inequality-} under an additional finite-trace constraint, a setting that has been recently studied independently in the literature. This reduction allows us to replace $\mathcal{L}_s$ by $1+\mathcal{L}_s$ in the compactness argument, which is particularly useful since $(-\Delta)^{s'} \lesssim 1+\mathcal{L}_s$ for all $0<s'<s$ (see \cite[Theorem 1.2]{F2009}). We then derive suitable uniform bounds to remove the finite-trace constraint. This step relies crucially on a new CLR-type inequality for the Hardy operator, established by Frank and the authors in \cite{DFTNT2024}. Moreover, unlike the Lieb--Thirring problem \eqref{LT-inequality}, the analysis of the Hardy--Lieb--Thirring problem \eqref{HLT-inequality-} requires additional information on the decay properties of the density of the optimizers. This delicate requirement is achieved by a detailed division into the radial and non-radial cases, each of which is of independent interest. \medskip

\noindent \textbf{Organization of the paper.} Theorem \ref{thm:optimizer-LT} and Theorem \ref{thm:optfinite} will be proved in Section \ref{sec:LT} and  Section \ref{sec:HLT} respectively. In the Appendix, concentration lemmas and IMS formulas will be established. \medskip
	
\noindent \textbf{Notations and conventions.} In the rest of the paper, for simplicity, we write $\| \cdot\|_{L^p}$ for 	$\| \cdot\|_{L^p(\R^d)}$. For $R>0$, we denote by $B_R$ the ball of radius $R$ and center at the origin. For an operator $A$, we denote by $\cN(0,A)$ the number of nonpositive eigenvalues of $A$. \medskip

\noindent \textbf{Acknowledgments.} Phan Th\`anh Nam would like to thank Yujin Guo for helpful exchanges concerning the parallel work \cite{CGWHLT2025}.  Giao Ky Duong and Phan Th\`anh Nam acknowledge partial support from the Deutsche
Forschungsgemeinschaft (DFG, German Research Foundation) through Germany’s Excellence Strategy EXC - 2111 - 390814868 and through TRR 352 – Project-ID 470903074. Thi Minh Thao Le and Phuoc-Tai Nguyen were supported by the Czech Science Foundation Project GA22-17403S. Part of the work was conducted during the visit of Thi Minh Thao Le and Phuoc-Tai Nguyen to the Department of Mathematics, LMU Munich, whose warm hospitality is gratefully acknowledged.

	\section{Proof of Theorem \ref{thm:optimizer-LT}} \label{sec:LT}
	
	\subsection{Existence of optimizers for the restricted problem} 
	
	Let 
		\begin{equation}\label{eq:LT-para}
		d \in \mathbb{N}, \quad s \in (0,1],\quad s<d/2,\quad q \in (\frac{d+2s}{d},\frac{d}{d-2s}),\quad \theta = \frac{d(q-1)}{2sq} \in (0,1).
			\end{equation}
	Recall that the sharp constant $C_{\rm LT}$ in \eqref{LT-inequality} is defined by
	$$ C_{\rm LT}= \inf_{\gamma \in \cK}\frac{ \| \gamma\|_{\dot{\cH}^s}^\theta}{\left\| \rho_\gamma \right\|_{L^q}}
	$$ 
	where 
	\begin{equation}\label{defi-K_1}
		\cK = \{ \gamma \text{ is self-adjoint such that } 0 \le \gamma \le 1\text{ and } \|\gamma\|_{\dot{\mathcal{H}}^s}  < +\infty  \}.
	\end{equation} 
	For $N \in \mathbb{N}$, we introduce the problem
	\begin{equation}\label{variation-N}
		\alpha_N =\inf_{  \substack{ \gamma \in \cK \\
				\tr(\gamma) \le N}} \frac{ \| \gamma\|_{\dot{\cH}^s}^\theta}{\left\| \rho_\gamma \right\|_{L^q}}  \in [C_{\rm LT}, +\infty).
	\end{equation}
	
	Note that the trace constraint $\tr(\gamma) \le N$ is ``linear", which is crucial for ensuring the existence of optimizers in \eqref{variation-N}. At first sight, this does not necessarily imply that the optimizer is of finite rank, but we will show that this is indeed the case. 
	
	Let us start with 

	\begin{lemma}[Existence of optimizers]\label{lem-exist-finite-trace-op-Lapace}  Under the parameters in \eqref{eq:LT-para}, there exists a minimizer $\gamma_N$ for the variational problem \eqref{variation-N}.     
	\end{lemma}
	\begin{proof}
		Let   $\{ \gamma_n\}_{n \ge 1}$ be  a minimizing sequence  for the  problem \eqref{variation-N}. By replacing $\gamma_n$ by $U_\lambda \gamma_n U_\lambda^*$ if necessary, where $U_\lambda$ is the unitary  $U_\lambda$ on $L^2(\R^d)$ defined,  for $\lambda >0$, by 
		\begin{equation}\label{define-U-lambda}
			[U_\lambda f](x)  = \lambda^\frac{d}{2} f(\lambda x), \text{ for } f \in L^2(\R^d),     
		\end{equation}
		we may assume that 
		\begin{equation}\label{minimizing-sequence-Laplace}
			0\le \gamma_n\le 1, \; \tr\gamma_n = \int_{\R^d}\rho_{\gamma_n} \leq N,\;
			\int_{\R^d} \rho_{\gamma_n}^q  = 1, 
			\; \tr ((-\Delta)^s \gamma_n)= \|\gamma_n\|_{\dot{\cH^s}}     \to  {\alpha}_N^{\frac{1}{\theta}}. 
		\end{equation}
		Since $\{\Tr \gamma_n\}_{n \geq 1}$ and $\{\Tr ((1  + (-\Delta)^s)\gamma_n)\}_{n\geq 1}$ are uniformly bounded, by the Banach--Alaoglu theorem and noting that $\left(1 + (-\Delta)^s\right)^{-1}$ is bounded in $L^2\left(\R^2\right)$, up to a subsequence of $\{\gamma_{n}\}_{n\geq 1}$, 	
		\begin{align}
			&	  (1  + (-\Delta)^s)^{\frac{1}{2}} \gamma_n (1  + (-\Delta)^s)^{\frac{1}{2}}
			\overset{\ast}{\rightharpoonup}
			(1  + (-\Delta)^s)^{\frac{1}{2}}\gamma_\infty (1  + (-\Delta)^s)^{\frac{1}{2}} 
			\label{converge-star-trace-class}\\
			&  \text{ and }  \gamma_n \overset{\ast}{\rightharpoonup} \gamma_\infty, \text{ as } n \to +\infty,   \text{(weakly-*) in trace class}. \nonumber
		\end{align}  	
		Consequently, we have { $ (-\Delta)^{s/2} \gamma_n (-\Delta)^{s/2} 
			\overset{\ast}{\rightharpoonup}
			(-\Delta)^{s/2} \gamma_\infty(-\Delta)^{s/2} $, which implies}
		$$
		0 \le \gamma_\infty \le 1, 
		\quad 
		\Tr \gamma_\infty \le N, 
		\quad 
		\Tr ((-\Delta)^s \gamma_\infty) \le  {\alpha}_N^{\frac{1}{\theta}}. 
		$$
		Thus, to conclude that $\gamma_\infty$ is an optimizer, the key point is to show that $\int_{\R^d} \rho_{\gamma_\infty}^{q}dx=1$. 
		
		\medskip
		\noindent \textbf{Step 1.} We are going to prove that for all $\chi \in C^\infty_c\left(\R^d\right)$, 
		\begin{equation}\label{storng-convergence-in-trace-class-laplace}
			\chi \gamma_n \chi \to   \chi \gamma_\infty \chi \quad \text{ strongly in trace class.}
		\end{equation}
		Indeed, for $\chi \in C^\infty_c\left(\R^d\right)$, we have $\chi \gamma_n\chi \overset{\ast}{\rightharpoonup} \chi \gamma_\infty \chi$ weakly-* in trace class and 
		\begin{align*}
			\Tr (\chi \gamma_n \chi)
			= 
			\Tr (\chi (1 + (-\Delta)^{s})^{-\frac{1}{2}} (1 + (-\Delta)^{s})^{\frac{1}{2}}\gamma_n (1 + (-\Delta)^{s})^{\frac{1}{2}} (1 + (-\Delta)^{s})^{-\frac{1}{2}} \chi  )\to 		\Tr (\chi \gamma_\infty \chi).
		\end{align*}
		Here we used \eqref{converge-star-trace-class} and the fact that $\chi (1 + (-\Delta)^{s})^{-1}\chi$ is a compact operator by Sobolev's embedding.  Therefore, \eqref{storng-convergence-in-trace-class-laplace} follows from the reciprocal of Fatou’s lemma (see e.g. \cite[Corollary 1]{Rob70}  or  \cite[Appendix H]{BS1979}). 
		
		From \eqref{storng-convergence-in-trace-class-laplace}, we also deduce that  
		$$\chi^2 \rho_{\gamma_n} = \rho_{\chi \gamma_n \chi} \to \rho_{\chi \gamma_\infty \chi} = \chi^2 \rho_{\gamma_\infty} $$
		strongly in $L^1(\R^d)$. Since $\chi\in C_c^\infty(\R^d)$ can be chosen arbitrarily, we obtain that, up to a subsequence,  $\rho_{\gamma_n}\to \rho_{\gamma}$ a.e. in $\R^d$.
		
		\medskip
        
		\noindent \textbf{Step 2} (No vanishing).  By the Lieb--Thirring inequality  \eqref{LT-inequality} and \eqref{minimizing-sequence-Laplace},  we find that  $\|\rho_{\gamma_n}\|_{L^{1+2s/d}} \lesssim 1$. Combining this with $\|\rho_{\gamma_n}\|_{L^1}\le N$ and $\| \rho_{\gamma_n}\|_{L^q}=1$, we conclude by the  \textit{pqr}-Lemma (see  \cite[Lemma 3.2]{FCIRM13}) that  the vanishing case in Lemma \ref{concentration-compactness-lemma} cannot occurs for the sequence  $\{\rho_{\gamma_n}^q\}_{n \geq 1}$.	Consequently, either the tightness for $\{\rho_{\gamma_n}^q\}_{n \geq 1}$ as in \eqref{concentrated-sequence} or  the  dichotomy as in \eqref{dichotomy-sequence} holds.
		
		\medskip
		\noindent \textbf{Step 3} (No dichotomy).  We prove that  the dichotomy scenario as in Lemma  \ref{concentration-compactness-lemma} can be ruled out. Suppose by contradiction that, up to a subsequence and a translation, there exist a sequence $\{R_{n}\}$ satisfying $R_n \nearrow \infty$ as $n \to \infty$ and a constant $\alpha \in (0,1)$   such that   
		\begin{equation} \label{eq:dichotomy}
			\int_{\R^d} \chi_n^{2q} \rho_{\gamma_n}^q dx= 1-\alpha + o(1)_{n\to \infty},
			\quad
			\int_{\R^d} \eta_n^{2q} \rho_{\gamma_n}^qdx = \alpha+ o(1)_{n\to \infty}, 
		\end{equation}
		and 
		$$ \int_{\{R_n  \le  |x| \le 4R_n\}} \rho_{\gamma_n}^q dx = o(1)_{n\to \infty},$$
		where $\chi_n$ and $\eta_n$ are defined in Appendix \ref{apendix:IMS} such that $\chi_n^2+\eta_n^2=1$.  By the localization formula in Lemma \ref{lemma-IMS}, we can decompose 
		\begin{align}\label{eq:localization}
			\|\gamma_n\|_{\dot{\cH^s}}  
			&
			= \|\chi_n\gamma_n\chi_n\|_{\dot{\cH}^s} + \|\eta_n\gamma_n\eta_n\|_{\dot{\cH}^s} + o(1)_{n\to \infty}\nonumber\\
			&		\ge 
			{\alpha}_N^{\frac{1}{\theta}}  \left(  \int_{\R^d} \rho_{\chi_n \gamma_n \chi_n}^q  dx \right)^{\frac{1}{q\theta}} 
			+ 
			{\alpha}_N^{\frac{1}{\theta}}   \left(  \int_{\R^d} \rho_{\eta_n \gamma_n \eta_n}^q  dx \right)^{\frac{1}{q\theta}}  + o(1)_{n\to \infty}.
		\end{align}
		Since $0<\frac{1}{q\theta}<1$, we have the inequality
		$$
		a^{\frac{1}{q\theta}} + b^{\frac{1}{q\theta}} \ge (1+\epsilon_\delta) (a+b) ^{\frac{1}{q\theta}} 
		$$
		for all $1\ge a,b \ge \delta>0$ and for some constant $\epsilon_\delta>0$ depending only on $\delta$. Consequently, thanks to \eqref{eq:dichotomy}, we deduce from \eqref{eq:localization} that 
		\begin{align*} \|\gamma_n\|_{\dot{\cH^s}}  	&\geq
			(1 + \epsilon_\alpha) {\alpha}_N^{\frac{1}{\theta}}
			\left( \int_{\R^d} \rho_{\chi_n \gamma_n \chi_n}^q dx +  \int_{\R^d} \rho_{\eta_n \gamma_n \eta_n}^q dx \right)^\frac{1}{q\theta} + o(1)_{n\to \infty} \nonumber\\
			& 
			=  (1 + \epsilon_\alpha){\alpha}_N^{\frac{1}{\theta}}.    
			\left( \int_{\R^d} \rho_{\gamma_n }^q \left(\chi_n^{2q} + \eta_n^{2q} \right)dx   \right)^\frac{1}{q\theta} +o(1)_{n\to \infty}\nonumber\\			
			&
			\ge (1 + \epsilon_\alpha){\alpha}_N^{\frac{1}{\theta}}+o(1)_{n\to \infty}. 
		\end{align*}
		The latter bound violates the minimizing property $\|\gamma_n\|_{\dot{\cH^s}}  \to {\alpha}_N^{\frac{1}{\theta}}$. Thus the dichotomy case cannot occur. 
		
		\medskip
		\noindent \textbf{Step 4} (Existence of minimizers). We infer from the previous steps that the sequence $\{\rho_{\gamma_n}^q\}_{n \geq 1}$ is tight. Combining this with the fact that $\rho_{\gamma_n}\to \rho_{\gamma_\infty}$ a.e. in $\R^d$ and that $\{ \rho_{\gamma_n}\}_{n \geq 1}$ is uniformly bounded in $L^{1+2s/d}(\R^d)$, as well as the assumption $q > 1+2s/d$, we conclude by using Lemma \ref{concentration-implies-strong-convergence} that $|\rho_{\gamma_n}|^q \to |\rho_{\gamma_\infty}|^q$ strongly in $L^1(\R^d)$. Thus $\int_{\R^d} \rho_{\gamma_\infty}^qdx=1$, and hence $\gamma_\infty$ is an optimizer for \eqref{variation-N}. Thus the proof is complete with the change of notation from $\gamma_\infty$ to $\gamma_N$.
	\end{proof}

	\subsection{Structure of the minimizer}
	In this subsection, we are going to derive the Euler Lagrange equations of  the optimizers.   
	\begin{lemma}[Spectral decomposition]\label{lem-existe_optimizer-N}  The minimizer $\gamma_N$ obtained in Lemma \ref{lem-exist-finite-trace-op-Lapace} can be decomposed as
		$$ \gamma_N =   \sum_{j= 1}^{J_N}  | \phi_{N,j}\rangle \langle  \phi_{N,j}|, $$ 
		where $J_N \in \N $ and $\{ \phi_{N,j}\}_{j=1}^{J_N} \subset H^s(\R^d)$ is an orthogonal sequence in $L^2(\R^d)$ with $\| \phi_{N,j} \|_{L^2} \leq 1$ for all $1 \leq j \leq J_N$.   Moreover, $\phi_{N,j}$ satisfies the Euler-Lagrange equations
		\begin{equation} \label{eigen} ((-\Delta)^s  - \rho_{\gamma_N}^{q-1})\phi_{N,j} = -\mu_{N,j}\phi_{N,j}, \quad j=1,2,\ldots,J_N,
		\end{equation}
		with $-\mu_{N,1} \leq -\mu_{N,2} \leq \ldots - \mu_{N,J_N} \leq 0$ denoting the lowest eigenvalues (counting multiplicities) of the operator $(-\Delta)^s - \rho^{q-1}_{\gamma_\infty}$.
	\end{lemma}
\begin{remark}
We note that the constraint $\tr(\gamma) \leq N$ in the problem \eqref{variation-N} is hidden in the non-positivity of the eigenvalue $\mu_{N,j}$ in the Euler-Lagrange equation \eqref{eigen} for $\phi_{N,j}$. 
\end{remark}
	
	The proof of Lemma \ref{lem-existe_optimizer-N} is divided into several steps presented below. 
	
		Applying the scaling transformation 
		$\gamma_N \mapsto \gamma_{N,\lambda} = U_\lambda \gamma_N U_\lambda^*$ if necessary, 
		where $\lambda = \left(\frac{\alpha_{N}}{\theta}\right)^{d-\frac{d}{q\theta}}$ and $U_\lambda$ is defined in \eqref{define-U-lambda}, 
		we may assume that $\gamma_N$ satisfies $\|\gamma_N\|_{\dot{\cH}^s} = \theta \|\rho_{\gamma_N}\|_{L^q}^q$.
	
		Now, we define  $V_N =\rho_{\gamma_N}^{q-1}$ and we     consider the   following variational problem
		\begin{equation} \label{vari-I_V-Hardy}
			I_{V_N; \text {min}}=\inf_{\gamma \in \mathcal{K}_1, \tr(\gamma) \le N} I_{V_N}(\gamma),
		\end{equation}
		where $I_{V_N}(\gamma) := \| \gamma\|_{\dot{\mathcal H}^s} - \int_{\R^d}V_N\rho_{\gamma}dx $.
		
		\medskip 
		\noindent
		\textbf{Step 1:} In  this step, we prove that $\gamma_N$ is also a minimizer for the problem   \eqref{vari-I_V-Hardy}. This is stated in the following lemma
		\begin{lemma} \label{lmm:mimimizerI-hardy} There holds    $I_{V_N;\text{\rm min}} = I_{V_N}\left(\gamma_N\right)$.
		\end{lemma}
		This result extends \cite[Lemma 13]{HKY2019} in which  the authors   treated the case $s=1$.  
		
		\begin{proof}[Proof of Lemma  \ref{lmm:mimimizerI-hardy}]	 Define the Weinstein functional  $W$ by
		$$
			W(\gamma) = \frac{ \| \rho_\gamma \|^q_{L^q}}{\| \gamma\|_{\dot{\cH}^s}^{q\theta} }, \quad \gamma \in \cK_N,
		$$
where 
	$$ \cK_N :=  \{  \gamma \in \cK: \,\tr(\gamma) \le N \}.
	$$	 
Since   $\gamma_N$ is a minimizer for the problem \eqref{variation-N},  it maximizes $W$  over $\cK_N$. Let  $\gamma \in \cK_N$ and  for $t \in [0,1]$, put $\gamma_t =\left(1-t\right)\gamma_N + t\gamma \in \cK_N$.    

For any $t \in [0,1]$, we have
			\begin{equation*}
				\begin{aligned} 
					0 
					\geq &
					\frac{W(\gamma_t)-  W(\gamma_N)}{t} 
					\\
					&=
					\frac{\|\rho_{ \gamma_t}\|_{L^q}^q-\|\rho_{\gamma_N}\|_{L^q}^q}{t} \cdot \frac{1}{\|\gamma_t\|_{\dot{\cH^s}}^{q \theta}} 
					+ \frac{\|\rho_{\gamma_N}\|_{L^q}^q}{t}\left(\frac{1}{\| \gamma_t\|_{\dot{\cH^s}}^{q \theta}}-\frac{1}{\|\gamma_N\|_{\dot{\cH^s}}^{q \theta}}\right).
				\end{aligned}
			\end{equation*}
			Letting $t \to 0^+$ yields
			\begin{equation*}
				\begin{aligned}
					0	
					\geq
					&
					\frac{d}{d t} \Big\vert_{t=0}\|\rho_{(1-t) \gamma_N +t \gamma}\|_{L^q}^q \cdot
					\frac{1}{
						\|\gamma_N\|_{\dot{\cH^s}}^{q \theta}} -
					\frac{q \theta \|\rho_{\gamma_N}\|_{L^q}^q}{\|\gamma_N\|_{\dot{\cH^s}}^{q \theta+1}}
					\cdot
					\frac{d}{d t}\Big\vert_{t=0}\|(1-t) \gamma_N + t \gamma\|_{\dot{\cH^s}} 
					\\
					&
					=
					\frac{ W(\gamma_N)}{\|\rho_{\gamma_N}\|_{L^q}^q} \cdot \frac{d}{d t}\Big \vert_{t=0}\left\|\rho_{(1-t) \gamma_N+t \gamma}\right\|_{L^q}^q-q \theta \frac{ W(\gamma_N)}{\|\gamma_N\|_{\dot{\cH}^s}} 
					\cdot 
					\frac{d}{d t}\Big\vert_{t=0}\|(1-t) \gamma_N + t \gamma\|_{\dot{\cH}^s}
					\\
					&
					=
					\frac{q   W(\gamma_N)}{\|\rho_{\gamma_N}\|_{L^q}^q}
					\left(\int_{\R^d} \rho_{\gamma_N}^{q-1}\left(\rho_\gamma-\rho_{\gamma_N}\right) d x-
					\left(\|\gamma\|_{\dot{\cH^s}} - \|\gamma_N\|_{\dot{\cH^s}} \right)
					\right).
				\end{aligned}
			\end{equation*}
			In the last equality, we have used the relation $\|\gamma_N\|_{\dot{\cH}^s} = \theta \|\rho_{\gamma_N}\|_{L^q}^q$ and the identities
			\begin{align*}
				&\frac{d}{d t}\Big\vert_{t=0}\|\rho_{(1-t) \gamma_N + t \gamma}\|_{L^q}^q = \frac{d}{d t}\Big\vert_{t=0}\|\rho_{\gamma_N} + t(\rho_{\gamma} - \rho_{\gamma_N})\|_{L^q}^q = 	q\int_{\R^d} \rho_{\gamma_N}^{q-1}\left(\rho_\gamma-\rho_{\gamma_N}\right) d x, \\ 
				&\frac{d}{d t}  \|(1-t) \gamma_N + t \gamma\|_{\dot{\cH}^s} 
				= 
				\frac{d}{dt} \left(\| \gamma_N\|_{\dot{\cH}^s} 
				+ t\| \gamma-\gamma_N\|_{\dot{\cH}^s}  \right) =  \| \gamma-\gamma_N\|_{\dot{\cH}^s} = \| \gamma\|_{\dot{\cH}^s} - \|\gamma_N\|_{\dot{\cH}^s}.
			\end{align*}
			Therefore,  we arrive at 
			\begin{equation*}
				I_{V_N}(\gamma_N)
				= \| \gamma_N\|_{\dot{\cH}^s}
				 -  \int_{\R^d} V_N \rho_{\gamma_N} dx
				\leq
				\| \gamma\|_{\dot{\cH}^s} -   \int_{\R^d} V_N \rho_{\gamma}dx = I_{V_N}(\gamma).
			\end{equation*}
			 As $\gamma$ is taken  arbitrarily, it follows  that  $\gamma_N$ is a minimizer  to  \eqref{vari-I_V-Hardy}. The proof of Lemma  \ref{lmm:mimimizerI-hardy} is complete.
		\end{proof}

		\medskip 
		\noindent 
		\textbf{Step 2:} In this step, let us consider the operator $(-\Delta)^s - V_N$ with $V_N= \rho_{\gamma_N}^{q-1}$. 
Since $\rho_{\gamma_N} \in L^r(\R^d)$ for all $r \in \bigl[1,\,\tfrac{d}{d-2s}\bigr)$, it follows that $V_N \in L^p(\R^d)$ for some $p > \frac{d}{2s}$. Therefore, by the Sobolev inequality, $V_N$ is form-bounded with respect to $(-\Delta)^s$ with a relative bound $\alpha \in (0,1)$. Hence, by the KLMN theorem \cite[Theorem X.17]{RS1975}, the operator $(-\Delta)^s - V_N$ admits a unique self-adjoint extension via the Friedrichs extension.  Moreover, 
$$
\sigma_{\mathrm{ess }}\,((-\Delta)^s - V_N) \;=\; [0,\infty).
$$

	Let $\{-\mu_{N,j}\}_{j=1}^{J^-_N}$ be the set of negative eigenvalues (counting multiplicities) of $(-\Delta)^s - V_N$ with the ordering 
	$$ -\mu_{N,1} \leq - \mu_{N,2} \leq \ldots < 0.
	$$
	Here  $J^-_N $ is finite thanks  to  the fractional CLR inequality and the fact  that $V_N \in L^\frac{d}{2s}(\R^d) $.  Let $\phi_{N,j}^-$ ($1 \leq j \leq J_N^-$) be the $L^2$-normalized eigenfunction corresponding to the eigenvalue $-\mu_{N,j}$, then  it  is a $H^s$-weak solution to
	$$
		((-\Delta)^s - V_N)\phi_{N,j}^- = -\mu_{N,j} \phi_{N,j}^-.
	$$
	We observe that if  $\gamma$ is a smooth finite-rank operator then
	\begin{equation*}
		I_{V_N}(\gamma)
		= 	\tr [((-\Delta)^s - V_N) \gamma].
	\end{equation*} 
	Thus, in order for $I_{V_N}$ to achieve its minimum value, $\gamma$ should contain as many as possible  $|\phi_{N,j}^-\rangle\langle\phi_{N,j}^-|$'s to be more negative, but its spectrum should not include any positive spectrum  of $(-\Delta)^s - V_N$.

	Let us introduce the projector
	\begin{equation} \label{eq:pinegative}
		\Pi^- := \sum_{j=1}^{J_N^-}|\phi_{N,j}^-\rangle\langle\phi_{N,j}^-|.
	\end{equation}
	The following lemma asserts that $\Pi^-$ is a	minimizer for \eqref{vari-I_V-Hardy}. It also confirms that $\Pi^- + \gamma^0_N$ is a minimizer for \eqref{vari-I_V-Hardy}, where   $\gamma^0_N$ is the part of $\gamma_N$ associated to the zero-eigenspace of  $(-\Delta)^s - V_N$.
	\begin{lemma} \label{lmm:decompImin-Hardy} For $N$ large enough, $\gamma_N$ can be decomposed as 
		$$\gamma_{N} = \Pi^- + \gamma^0_N, $$
		where $\Pi^-$ is given in \eqref{eq:pinegative} and $\gamma^0_N$ is a self-adjoint operator acting on the eigenspace associated with the zero eigenvalue of the operator  $(-\Delta)^s - V_N$.   In other words,  $\gamma^0_N$ can be zero or admits the following form
		$$   \gamma^0_N  =   \sum_{j=1}^{J_N^0}|\phi_{N,j}^0\rangle\langle\phi_{N,j}^0|,$$
		where $\phi_{N,j}^0 \in \ker((-\Delta)^s - V_N) $ and $J_N^0$ is finite. 
	\end{lemma}
	\begin{proof}[Proof of Lemma \ref{lmm:decompImin-Hardy}]
		We  have $\Pi^- \in  \cK_N $ for $N$ large enough since 
		$$  \tr(\Pi^-)  =   \ran(\Pi^-) \le  \cN(0,(-\Delta)^s - V_N) \lesssim_{d,s}   \int_{\R^d} V_N^{\frac{d}{2s}}dx \lesssim_{d,s,q} 1,   $$
		where $ \cN(0,(-\Delta)^s - V_N)$ denotes the number of negative and zero  eigenvalues of $(-\Delta)^s - V_N$ (counting multiplicities). 
		Given any arbitrary  smooth finite-rank operator $ \gamma \in \cK_N$, we  write
		\begin{equation} \label{IV(gamma)}
			I_{V_N}(\gamma)=I_{V_N}(\Pi^{-})+I_{V_N}(\bar \gamma), \quad \text{where } \bar \gamma  =  \gamma -\Pi^-.
		\end{equation}
		In order to show that $\Pi^-$ is a minimizer for \eqref{vari-I_V-Hardy}, it is sufficient to prove 
		\begin{equation} \label{est:tildegamma}
			I_{V_N}(\bar \gamma) \geq 0.
		\end{equation}
		By  the linearity, we need to prove \eqref{est:tildegamma} when $\bar \gamma$ is a one-particle projection of the form either $-|\phi\rangle\langle\phi|$ or $|\psi\rangle\langle\psi|$. 
		Since $-\Pi^- \leq \bar \gamma \leq 1 -\Pi^-$, $\phi$ must be contained in ${\rm span}\{\phi^-_{N,j}\}_{j= 1}^{J_N^-}$ and $\psi$ must be orthogonal to ${\rm span}\{\phi^-_{N,j}\}_{j= 1}^{J_N^-}$. Thus, we arrive at
		$$
			I_{V_N} (-|\phi\rangle\langle\phi|)
			=-\|   (-\Delta)^\frac{s}{2}\phi\|_{L^2}^2+\int_{\R^d} V_N|\phi|^2 dx>0
		$$
		and
		$$
			I_{V_N}(|\psi\rangle\langle\psi|)
			=
			\|(-\Delta)^\frac{s}{2} \psi\|_{L^2}^2-\int_{\R^d} V_N|\psi|^2 d x \geq 0 ,
		$$
		which leads to \eqref{est:tildegamma}. Therefore, we infer from \eqref{IV(gamma)} that $I_{V_N}(\Pi^-) \leq I_{V_N}(\gamma)$. By the density of finite-rank operators in $\cK_N$ we derive that $I_{V_N}(\Pi^-) \leq I_{V_N,\rm min}$, which means that  $\Pi^-$ is a minimizer of \eqref{vari-I_V-Hardy}.
		
		Put $\gamma^0_N  = \gamma_{N} - \Pi^-$, then we have 
		\begin{equation*}
			I_{V_N; \min }
			=
			I_{V_N}(\gamma_{N })
			=
			I_{V_N}(\Pi^{-}) + I_{V_N}(\gamma_N^0)
			=
			I_{V_N; \min }+I_{V_N}(\gamma_N^0),
		\end{equation*}
		which yields
		\begin{equation*}
			I_{V_N}(\gamma_N^0) = \tr[((-\Delta)^s  -V_N)\gamma_N^0]= 0.
		\end{equation*}
		In addition, since $\gamma_N^0$ belongs to the orthogonal complement of the negative eigenspace of $(-\Delta)^s -V_N$, it must lie entirely within the eigenspace corresponding to the zero eigenvalue of $(-\Delta)^s -V_N$. In particular, $\gamma_N^0$ can be written in the form 
		$$ \gamma_N^0 =     \sum_{j=1}^{J_N^0}|\phi_{N,j}^0\rangle\langle\phi_{N,j}^0|,$$
		where $  \phi_{N,j}^0  \in \ker((-\Delta)^s -V_N)$. Note that, from fractional CLR inequality and the fact  $V_N \in L^\frac{d}{2s}(\R^d) $, we derive that $J_N^0 < \infty$. Moreover, by the orthogonality and the fact that $0 \leq \gamma_N \leq 1$, we have  $\|\phi_{N,j}^0\|_{L^2} \le 1$ for any $j =1,2,\ldots,J_N^0$. The proof of Lemma \ref{lmm:decompImin-Hardy} is complete.
	\end{proof}
	
	From  Step 1 and Step 2, we get the desired results in Lemma \ref{lem-existe_optimizer-N}.

\subsection{Conclusion of Theorem \ref{thm:optimizer-LT}}\label{section-conclu-Theo-1}

The key ingredient of the proof of Theorem \ref{thm:optimizer-LT} is the following uniform estimate. 
\begin{lemma}\label{lem_staionary-seq}    Let $\gamma_N$ be the minimizer for  the restricted problem \eqref{variation-N} obtained in Lemma \ref{lem-exist-finite-trace-op-Lapace}.   Then,  for all $N \ge 1$,
\begin{equation} \label{Tr < ran}  \Tr (\gamma_N)\le \ran(\gamma_N) \le N_0, 
\end{equation}
	where the constant $N_0 = N_0(d,s,q) \in \mathbb{N}$ is independent of $N$.
\end{lemma}
\begin{proof}[Proof of Lemma \ref{lem_staionary-seq}] By Lemma \ref{lem-existe_optimizer-N}, we have 
\begin{equation} \label{Tr} \tr(\gamma_N) \leq \ran(\gamma_N) \leq \cN(0,(-\Delta)^s - V_N)
\end{equation}
where $V_N = \rho_{\gamma_N}^{q-1} $.  This, in combination with the CLR inequality \cite{Rozen72CLR,Lieb1976CLR,Cwikel77}  (see also, e.g., \cite{Frank20,Nam-review} for recent reviews, including the fractional case) and H\"older's inequality, implies that, for any $\alpha \in (0,1)$,
	$$  \ran(\gamma_N) \lesssim_{d,s}  \int_{\R^d} V_N^{\frac{d}{2s}} dx \le \left( \int_{\R^d}  \rho_{\gamma_N} dx \right)^{\alpha}  \left(  \int_{\R^d}  \rho_{\gamma_N}^{ \frac{{1}}{1- \alpha} \left( \frac{d}{2s}(q-1) - \alpha\right) } dx \right)^{1-\alpha}.   $$
	We can choose $\alpha \in (0,1)$ such that 
	$$ \frac{1}{1-\alpha} \left(   \frac{d}{2s}(q-1) - \alpha    \right) \in \left(   \frac{d + 2s}{d}, \frac{d}{d-2s}\right).$$
Note that $\|\rho_{\gamma_N}\|_{L^{1+2s/d}} = 1$ and $
\|\rho_{\gamma_N}\|_{L^{d/(d-2s)}} = (\alpha_N)^{1/\theta} \lesssim_{d,s,q} 1$ due to the fact that the sequence $\{\alpha_N\}_{N \geq 1}$ is decreasing.  
Consequently,
$$
\ran(\gamma_N) \lesssim_{d,s,q} \left(\int_{\mathbb{R}^d} \rho_{\gamma_N}\, dx \right)^{\alpha} 
\leq (\ran(\gamma_N))^{\alpha},
$$
which implies that $\operatorname{ran}(\gamma_N) \leq N_0$ for some $N_0=N_0(d,s,q)$ independent of $N$.   
This and \eqref{Tr} imply \eqref{Tr < ran}. The proof is complete.
\end{proof}

Now we are ready to conclude  Theorem \ref{thm:optimizer-LT}.
\begin{proof}[Proof of Theorem \ref{thm:optimizer-LT}] Let $N_0$ be the constant in Lemma \eqref{lem_staionary-seq}. We will show that $\gamma_{N_0}$ is an optimizer for \eqref{LT-inequality}.

From \eqref{variation-N} with $N=N_0$, we find that
$$ \frac{\| \gamma_{N_0}\|_{\dot{\cH}^s}^\theta}{\| \rho_{\gamma_{N_0}}\|_{L^q}} = \alpha_{N_0} \geq C_{\rm LT}.
$$	

To prove the reversed inequality, we will prove the following 

\begin{lemma}\label{lem_add} For any self-adjoint operator $0\le A \le 1$ on $L^2(\R^d)$ such that $(-\Delta)^{s/2} A (-\Delta)^{s/2}$ is trace class, we have
	\begin{align}\label{eq:LT-full-range-N0}
		\Tr( (-\Delta)^s A))^\theta \ge \alpha_{N_0} \|\rho_A\|_{L^q}. 
	\end{align}
	\end{lemma}
Note that here we do not assume that $A$ is trace class. 
	\begin{proof}[Proof of Lemma \ref{lem_add}]
	Indeed, let $0\le A \le 1$ such that $(-\Delta)^{s/2} A (-\Delta)^{s/2}$ is trace class, then we have the spectral decomposition 
	$$ A = \sum_{j\ge 1}  |\phi_{j}\rangle \langle \phi_{j}|$$
	where  $\{\phi_j\}_{j \geq 1}\subset H^s(\R^d)$ are orthogonal functions in $L^2(\R^d)$ with $\| \varphi_j \|_{L^2} \leq 1$ for all $j\geq 1$. Then for any $N_0 \le N\in \mathbb{N}$, by using \eqref{variation-N}, the fact that $\alpha_N=\alpha_{N_0}$ and Lemma \ref{lem_staionary-seq}, we have 
	$$
	\Tr( (-\Delta)^s A)) ^\theta \ge \left(  \sum_{j= 1}^N \| (-\Delta)^{\frac{s}{2}} \varphi_j\|^2_{L^2}\right)^\theta  \ge \alpha_N \|\rho_{A_N}\|_{L^q} = \alpha_{N_0} \|\rho_{A_N}\|_{L^q},
	$$
	where $A_N=\sum_{j =  1}^N  |\phi_{j}\rangle \langle \phi_{j}|$ and  $\rho_{A_N}(x)=\sum_{j=1}^N |\phi_j(x)|^2$. Since $\rho_{A_N}(x) \uparrow \rho_A(x)$ for a.e.  $x \in \R^d$ as $N \to \infty$, by the monotone convergence theorem, we obtain 
	$$
	\Tr( (-\Delta)^s A)) ^\theta \ge  \alpha_{N_0} \lim_{N\to \infty }  \|\rho_{A_N}\|_{L^q} = \alpha_{N_0} \|\rho_A\|_{L^q}.
	$$
	Thus we have proved the claim of Lemma \ref{lem_add}. 
	\end{proof}
	
	Now, thanks to Lemma \ref{lem_add}, we have $\alpha_{N_0} \leq C_{\rm LT}$. Therefore, $\gamma_{N_0}$ is a finite-rank optimizer for \eqref{LT-inequality}. The representation of the optimizer follows from Lemma \ref{lem-existe_optimizer-N}. 
	
	The nonexistence part of the Theorem follows from Lemma \ref{lem_staionary-seq}. The proof of Theorem \ref{thm:optimizer-LT} is complete.
\end{proof}

\section{Proof of Theorem \ref{thm:optfinite}} \label{sec:HLT}
This section is devoted to the proof of Theorem \ref{thm:optfinite}. 

\subsection{Hardy--Lieb--Thirring inequality}     
\begin{theorem}[Hardy--Lieb--Thirring inequality]\label{theorem-Optimizer-HLT}
	Let $d \ge 1$, $s \in (0,1]$, $s<\frac{d}{2}$, $q \in [\frac{d+2s}{d},\frac{d}{d-2s})$ and $\theta=\frac{d(q-1)}{2sq}$. 
	Then there exists a constant $C=C(d,s,q)>0$ such that 
	\begin{equation}\label{HLT-inequality}
		\|\gamma\|_{op}^{1 -\theta} \| \gamma\|_{\dot{\cQ}^s}^\theta \ge C	\| \rho_\gamma \|_{L^q}
	\end{equation}
for all self-adjoint, bounded non-negative operators $\gamma$ such that $\|\gamma\|_{\dot{\mathcal{Q}}^s}  < +\infty$.	
\end{theorem}
\begin{proof} We consider the case $s\in (0,1)$ since the case $s =1$ can be treated in a similar, even simpler, way.  Since $\gamma$ is bounded, \eqref{HLT-inequality} is equivalent to
	\begin{equation}\label{esti-rho-gamma-lq-les-normH-theta}
		\| \gamma\|_{\dot{\cQ}^s}^\theta \geq C\|\rho_\gamma\|_{L^q}, \quad \text{for all }  \gamma \in \tilde \cK,
	\end{equation}
	where 
	\begin{equation}\label{defi-Q-1}
		\tilde \cK := \{\gamma \text{ is self-adjoint such that }  0 \le \gamma  \le 1 \text{ and }  \|\gamma\|_{\dot{\mathcal{Q}}^s}  < +\infty  \}.
	\end{equation}
	As $ \sqrt{\mathcal{L}_s}  \gamma \sqrt{\mathcal{L}_s}$ is trace class, the following spectral decomposition holds
	$$
		\gamma = \sum_{j=1}^\infty    | \phi_j  \rangle \langle  \phi_j|,  
	$$
	where  $\{\phi_j\}_{j \geq 1} \subset Q^s$ are orthogonal functions in $L^2(\R^d)$ with $\| \phi_j \|_{L^2} \leq 1$ for all $j \in \N$. 
	
	On one hand, by the Hardy--Lieb--Thirring inequality (see \cite{FLSJAMS08}), we have 
	\begin{equation} \label{eq:HLTgamma}
		\|\gamma\|_{\dot{\cQ}^s}  = \Tr |\mathcal{L}_s \gamma| \,    
	 \gtrsim_{d,s}  \, \int_{\R^d} \rho_\gamma^{1 + \frac{2s}{d}}dx.
	\end{equation} 
	On the other hand, we infer from the Hoffmann--Ostenhof inequality (see \cite{HOPRA77}) that
	\begin{equation}\label{eq:HOgamma}
		\|\gamma\|_{\dot{\cQ}^s} 
		\ge  
		\left\langle \mathcal{L}_s \sqrt{\rho_\gamma}, \sqrt{\rho_\gamma}  \right\rangle. 
	\end{equation} 
	Combining \eqref{eq:HLTgamma} and \eqref{eq:HOgamma} yields
	\begin{equation*}
		\|\gamma\|_{\dot{\cQ}^s}
		\gtrsim
		\left(\int_{\R^d} \rho_\gamma^{1+\frac{2s}{d}}dx\right)^{1-\nu}
		\langle \mathcal{L}_s \sqrt{\rho_\gamma}, \sqrt{\rho_\gamma}  \rangle^{\nu},
	\end{equation*}
	where $\nu \in [0,1)$ will be chosen later. Therefore, in order to prove  \eqref{esti-rho-gamma-lq-les-normH-theta}, it is sufficient to show that
	\begin{equation} \label{eq:CKNgamma}
		\left(\int_{\R^d} \rho_\gamma^{1+\frac{2s}{d}}dx\right)^{\left(1-\nu\right)}
		\langle \mathcal{L}_s \sqrt{\rho_\gamma}, \sqrt{\rho_\gamma}  \rangle^{\nu}
		\gtrsim
		\left\|\rho_\gamma\right\|_{L^q}^{\frac{1}{\theta}}.
	\end{equation}
	
	By a simple density  argument, we may assume that $\rho_\gamma \in C^\infty_c(\R^d \backslash \{0\})$. Using the  ground state representation  (see \cite[Proposition 4.1]{FLSJAMS08})  $v(x) = |x|^{\frac{d-2s}{2}}\sqrt{\rho_\gamma(x)}$, we find that \eqref{eq:CKNgamma} is equivalent to  
	\begin{equation}\label{equivalent-inequality-}
		\left\| \left|x\right|^{-\frac{d-2s}{2}}v \right\|_{L^{2\left(1 + \frac{2s}{d} \right)}}^{\left(1 + \frac{2s}{d} \right)\left(1-\nu\right)\theta}
		\left(\int_{\R^d} \int_{\R^d}  \frac{|v(x) -v(y)|^2}{|x-y|^{d+2s}} \frac{d x}{\left|x\right|^{\frac{d-2s}{2}}} \frac{d y}{\left|y\right|^{\frac{d-2s}{2}}}\right)^{\frac{\nu\theta}{2}}
		\gtrsim
		\left\|\left|x\right|^{-\frac{d-2s}{2}}v\right\|_{L^{2q}}.
	\end{equation}
	Due to the scaling property of \eqref{equivalent-inequality-}, we choose
	\begin{equation} \label{nu} \nu = \frac{qd - (d+2s)}{2s(q-1)}.
	\end{equation}
	Applying the fractional  Caffarelli--Kohn--Nirenberg inequality (see   \cite[Theorem 1.1]{SNJFA18})  with the parameters
	\begin{equation*}
		\begin{aligned}
			&\alpha = \beta = \gamma = -\frac{d-2s}{2},
			\quad
			\textbf{a} = \frac{d\left(qd - d-2s\right)}{4qs^2},
			\\
			&{\tau} =2q, 
			\quad
			\textbf{p}=2, 
			\quad\textbf{q} = 2\left(1+\frac{2s}{d}\right),
		\end{aligned}
	\end{equation*} 
	we derive \eqref{equivalent-inequality-} with $\nu$ as in \eqref{nu}. Thus we  obtain \eqref{HLT-inequality}, hence the proof is complete.
\end{proof}

\subsection{Optimizers  for  the restricted problem}
Let $d \ge 1$, $s \in (0,1]$, $s<\frac{d}{2}$,  $q \in \left( \frac{d+ 2s}{d}, \frac{d}{d-2s}\right)$ and $\theta  =  \frac{d(q-1)}{2sq}$. Recall that  the sharp constant $C_{\rm HLT}$ in \eqref{HLT-inequality-} is defined by
$$ C_{\rm HLT} = \inf_{\gamma \in \tilde \cK}\frac{ \| \gamma\|_{\dot{\cQ}^s}^\theta}{\| \rho_\gamma\|_{L^q}} > 0,
$$
where $\tilde \cK$ is defined in \eqref{defi-Q-1}. For every $N \in \mathbb{N}$, we introduce the variational problem
\begin{equation}\label{variation-N-Hardy}
	\beta_N =\inf_{  \substack{ \gamma \in \tilde \cK \\
			\tr(\gamma) \le N}} \frac{ \| \gamma\|_{\dot{\cQ}^s}^\theta}{\| \rho_\gamma\|_{L^q}}  \in [C_{\rm HLT}, +\infty).
\end{equation}

\medskip
\noindent 
\textbf{Step 1: Existence of a finite-trace optimizer.} In this step, we prove the following result. 
\begin{lemma}[Existence of optimizers]\label{lem-exist-finite-trace-op-Hardy}  There exists a minimizer $\tilde \gamma_N$ for problem \eqref{variation-N-Hardy}.
\end{lemma} 
\begin{proof}
	Let   $\{ \gamma_n\}_{n \ge 1}$ be  a minimizing sequence  for the  problem \eqref{variation-N-Hardy}.  By scaling, we may assume that 
	\begin{equation}\label{minimizing-sequence-Laplace-}
		0\le \gamma_n\le 1, \; \tr\gamma_n = \int_{\R^d}\rho_{\gamma_n}dx \leq N,\;
		\int_{\R^d} \rho_{\gamma_n}^q dx  = 1, 
		\; \tr (\cL_s\gamma_n)= \|\gamma_n\|_{\dot{\cQ^s}}     \to  \beta_N^{\frac{1}{\theta}}. 
	\end{equation}
	Since  $\{\tr\gamma_n\}_{n\geq 1}$ and $\{\tr((1  + \mathcal{L}_s)\gamma_n)\}_{n \geq 1}$ are uniformly  bounded,  there exist a subsequence, still denoted by $\{\gamma_n\}$, and  $\tilde \gamma_\infty \in \tilde \cK$  satisfying
	$\|\tilde \gamma_\infty\|_{\dot{\cQ}^s}  \le  {\beta}_N^{\frac{1}{\theta}}$ and $\Tr \tilde \gamma_\infty \le N$ such that
	\begin{align*}
		&	  \left(1  + \mathcal{L}_s\right)^{\frac{1}{2}} \gamma_n \left(1  + \mathcal{L}_s\right)^{\frac{1}{2}}
		\overset{\ast}{\rightharpoonup}
		(1  + \mathcal{L}_s)^{\frac{1}{2}}\tilde \gamma_\infty (1  + \mathcal{L}_s)^{\frac{1}{2}} \\
		&  \text{ and }  \gamma_n \overset{\ast}{\rightharpoonup} \tilde \gamma_\infty, \text{ as } n \to +\infty,   \text{(weakly-*) in trace class}. \nonumber
	\end{align*}   	
	Since $ \int_{\R^d} \rho_{\gamma_n}^qdx = 1$ for all $n \in \N$,  in order to prove that  $\tilde \gamma_\infty$ is a minimizer for problem \eqref{variation-N-Hardy}, we need to show that 
	$$\int_{\R^d} \rho_{\tilde \gamma_\infty}^q dx= 1.$$
	
	\medskip
	\noindent 
	{\bf Claim 1:} There holds 
	\begin{equation} \label{convergence-rho_n}
		\rho_{\gamma_n}(x)\to \rho_{\gamma}(x) \text{ for  a.e. }  x\in \R^d. 
	\end{equation}

	Let $\chi \in C^\infty_c\left(\R^d\right)$ and $s' <s$,  then 
\begin{align*}
	\chi (1 + \mathcal{L}_s)^{-1} \chi = \chi (1 + (-\Delta)^{s'})^{-1}    (1 + (-\Delta)^{s'}) (1 + \mathcal{L}_s)^{-1} \chi
\end{align*}
is compact	due to  the facts that $\chi (1 + (-\Delta)^{s'})^{-1} $ is compact (see \cite[Theorem 4.1]{BS1979}) and $(1 + (-\Delta)^{s'}) (1 + \mathcal{L}_s)^{-1} \chi$ is bounded (see \cite[Theorem 1.2]{F2009}).
	By using an argument similar to the one leading to \eqref{storng-convergence-in-trace-class-laplace},  we deduce from  the identity
	$$
	\Tr (\chi \gamma_n \chi)= \Tr (\chi (1 + \mathcal{L}_s)^{-\frac{1}{2}} (1 + \mathcal{L}_s)^{\frac{1}{2}}\gamma_n (1 + \mathcal{L}_s)^{\frac{1}{2}} (1 + \mathcal{L}_s)^{\frac{1}{2}}   \chi   \to 		\Tr (\chi \tilde \gamma_\infty \chi),
	$$
	that $\chi \gamma_n \chi \to   \chi \gamma_\infty \chi$  strongly in trace class.  
	Since $\chi \in C_c^\infty(\R^d)$  is arbitrary, we derive  \eqref{convergence-rho_n}.  
	
	\medskip
	\noindent 
	{\bf Claim 2:} Neither the vanishing scenario nor  the dichotomy scenario occurs.  
	
	Indeed, by the Hardy--Lieb--Thirring inequality  \eqref{HLT-inequality} with $q = 1 + \frac{2s}{d}$ and  by arguing along the same line as in the proof of Step 2 in Lemma \ref{lem-exist-finite-trace-op-Lapace}, we conclude that either the sequence \(\{\rho_{\gamma_n}^q\}_{n \in \N}\) is tight as in \eqref{concentrated-sequence} or a dichotomy occurs as in \eqref{dichotomy-sequence}. However, the  dichotomy scenario is excluded by repeating the argument in Step 3  of Lemma \ref{lem-exist-finite-trace-op-Lapace} and using \eqref{IMS-Hardy} instead of \eqref{IMS-Laplace}. 
	
	Therefore, thanks to
	Lemma \ref{concentration-compactness-lemma}, we obtain
	$$ \rho_{\gamma_n} \to  \rho_{ \tilde\gamma_\infty} \text{ strongly in } L^q(\R^d). $$ 
	Note that the norm $\left\|\cdot\right\|_{\dot{\cQ}^s}$ is not invariant under translation, hence applying Lemma \ref{lmm:bounded-translation-Ls} implies that
	\begin{equation*}
		\lim\limits_{R \to \infty}\int_{B_R}\rho_{\gamma_n}^qdx = 1.
	\end{equation*}
	Thus, by applying Lemma \ref{concentration-implies-strong-convergence}, we have $|\rho_{\gamma_n}|^q \to |\rho_{\tilde \gamma_\infty}|^q$ strongly in $L^1(\R^d)$, which implies
		$$ \int_{\R^d} \rho_{\tilde \gamma_\infty}^q dx =1.$$
	Therefore $\tilde \gamma_{\infty}$ is a minimizer for problem \eqref{variation-N-Hardy}. Thus we conclude the proof by changing the notation from $\tilde \gamma_\infty$ to $\tilde \gamma_N$.
\end{proof}

    \medskip 
Let $\tilde \gamma_N$ be the optimizer for problem \eqref{variation-N-Hardy} obtained in Lemma \ref{lem-exist-finite-trace-op-Hardy}. Let us consider the Hardy--Schrödinger operator  
$\mathcal{L}_s - \widetilde V_N$ where $\widetilde V_N = \rho_{\tilde \gamma_N}^{q-1}$.  
By the improved Hardy inequality (see \cite[Theorem 1.2]{F2009}), we have  
$$
	\mathcal{L}_s + \ell^{-2s'} \;\ge\; C_{d,s,s'}\, \ell^{-2(s-s')} (-\Delta)^{s'}, 
	\qquad \ell>0,\quad 0<s'<s<\tfrac{d}{2},
$$
where the constant \(C_{d,s,s'}\) depends only on \(d,s,s'\).  From this estimate, it follows that for $s' < s$ sufficiently close to $s$, the potential $\widetilde V_N$
is form-bounded with respect to $(-\Delta)^s$ with a relative bound $\alpha' \in (0,1)$. Hence, by the KLMN theorem \cite[Theorem X.17]{RS1975}, the operator $\mathcal{L}_s - \widetilde V_N$ admits a unique self-adjoint extension.  
Furthermore, by combining Sobolev inequalities with Weyl’s theorem on the essential spectrum \cite[Corollary 2, p.113]{RS1978}, we obtain  
$$
\sigma_{\mathrm{ess}}(\mathcal{L}_s - \widetilde V_N) \;=\; \sigma_{\mathrm{ess}}(\mathcal{L}_s) \;=\; [0,+\infty).
$$ 
Therefore, by employing a similar argument as in the proof of  Lemma~\ref{lem-existe_optimizer-N}, we obtain the following result 
\begin{lemma}\label{lem-existe_optimizer-N-hardy}  Let  $\tilde \gamma_N$  be   the optimizer in Lemma  \ref{lem-exist-finite-trace-op-Hardy}. Then $\tilde \gamma_N$ can be decomposed as         
	$$ \tilde \gamma_N =   \sum_{j =1 }^{\tilde J_N}   | \tilde \phi_{N,j}\rangle \langle  \tilde \phi_{N,j}|,  $$ 
	where $\tilde J_N \in \N $ and  $\{ \tilde \phi_{N,j}\}_{j=1}^{\tilde J_N} \subset Q^s$ is an orthogonal sequence in $L^2(\R^d)$ with $\|\tilde \phi_{N,j}\|_{L^2} \le 1$ for all $1 \leq j \leq \tilde J_N$.  Moreover, $\tilde \phi_{N,j}$ satisfies the Euler-Lagrange equations
	\begin{equation*} (\mathcal{L}_s  - \rho_{\tilde \gamma_N}^{q-1}) \tilde \phi_{N,j} = -\tilde \mu_{N,j} \tilde \phi_{N,j}, \quad j=1,2,\ldots,\tilde J_N,
	\end{equation*}
	with $-\tilde \mu_{N,1} \leq -\tilde \mu_{N,2} \leq \ldots - \tilde \mu_{N,\tilde J_N} \leq 0$ denoting the lowest eigenvalues (counting multiplicities) of the operator $\cL_s - \rho^{q-1}_{\tilde \gamma_\infty}$.  
\end{lemma}

\medskip
\noindent 
\textbf{Step 2: Uniformly bounded trace.} 
\begin{lemma}\label{lem-stationary-beta_N}
	Let $\tilde \gamma_N$ be the minimizer for  the restricted problem \eqref{variation-N-Hardy}  obtained in Lemma \ref{lem-exist-finite-trace-op-Hardy}.   Then,  for any $N \ge 1$,
	\begin{equation} \label{est:trace-ran}  \Tr (\tilde \gamma_N)\le \ran(\tilde \gamma_N) \le N_1, 
	\end{equation}
	where the constant $N_1=N_1(d,s,q) \in \N$ is independent of $N$.
\end{lemma}
\begin{proof}
We infer from Lemma \ref{lem-existe_optimizer-N-hardy} that
$$ \tr(\tilde \gamma_N) \leq \ran(\tilde \gamma_N) \leq \cN(0,\cL_s - \widetilde V_N).
$$
where $\widetilde V_N = \rho_{\tilde \gamma_N}^{q-1}$. By a suitable  scaling, we may assume that 
	\begin{equation}  \label{est:trace(Lgamma)}
		\tr(\sqrt{\cL_s}\tilde \gamma_N \sqrt{\cL_s}) \lesssim_{d,s,q} 1. 
	\end{equation}
In addition, by the  Hardy--Lieb--Thirring inequality  \eqref{HLT-inequality}, it follows that  \begin{equation}\label{est:rho-Lp}
		\|\rho_{\tilde \gamma_N}\|_{L^r} \lesssim_{d,s,r} 1, \quad \text{ for any } r \in \left[\frac{d+2s}{d},\frac{d}{d-2s}\right).
	\end{equation}
	
	Denote by $P$  the projection onto radial functions and $P^{\perp} = 1- P$. By the Cauchy--Schwarz inequality,  we have 
	\begin{equation*}
		\widetilde V_N \leq 2P \widetilde V_NP + 2P^\perp\widetilde V_N P^\perp ,
	\end{equation*}
	which yields 
	\begin{equation*}
		\cN(0, \mathcal{L}_s - \widetilde V_N)
		\leq
		\cN(0,P( \mathcal{L}_s - 2\widetilde V_N)P)
		+
		\cN(0,P^\perp( \mathcal{L}_s - 2\widetilde V_N)P^\perp).
	\end{equation*}
	
	For the non-radial part, by  \cite[Theorem 1.5 and Corollary 1.7]{BM2024}, we have
	$$P^\perp( \mathcal{L}_s - 2 \widetilde V_N)P^\perp 
	\gtrsim
	P^\perp( (-\Delta)^s - 2 \widetilde V_N)P^\perp, $$
	which, together with the CLR inequality for fractional Laplacian, implies
	$$ \cN(0,P^\perp( \mathcal{L}_s - 2 \widetilde V_N)P^\perp) \lesssim_{d,s} \cN(0,P^\perp( (-\Delta)^s - 2 \widetilde V_N)P^\perp) \lesssim_{d,s} \int_{\R^d} (P^\perp \widetilde V_N P^\perp)^{\frac{d}{2s}}dx.
	$$
	
	For the radial part, using  the CLR inequality  in   \cite[Theorem 6]{DFTNT2024}, we have
	$$ \cN(0,P( \mathcal{L}_s - 2 \widetilde V_N)P) \lesssim_{d,s} 1 + \int_{\R^d} (P \widetilde V_N P)^{\frac{d}{2s}}(1 + |\ln|x||)^{\frac{d-s}{s}}dx.
	$$
	Combining the above estimates yields
	\begin{equation} \label{eq:m-CLR} \begin{aligned}
		\tr(\tilde \gamma_N) &\leq \ran(\tilde \gamma_N) \\
		&\lesssim_{d,s}
		1
		+
		\int_{\R^d}(P \widetilde V_N P)^{\frac{d}{2s}}(1 + |\ln|x||)^{\frac{d-s}{s}}dx
		+
		\int_{\R^d}(P^\perp \widetilde V_N P^\perp)^{\frac{d}{2s}}dx
		\\
		&
		\lesssim_{d,s,q} 1 + 
		\int_{B_1}(P \widetilde V_N P)^{\frac{d}{2s}}|\ln|x||^{\frac{d-s}{s}}dx
		+
		\int_{B_1^c}(P \widetilde V_N P)^{\frac{d}{2s}}|\ln|x||^{\frac{d-s}{s}}dx \\
		&\quad +
		\int_{\R^d}(P^\perp \widetilde V_N P^\perp)^{\frac{d}{2s}}dx.
	\end{aligned} \end{equation}

In order to control the  terms on the right hand side of \eqref{eq:m-CLR},  we  treat the following cases separately: 
	$$
	q \in \left( \frac{d^2 + 2ds + 4s^2}{d^2}, \, \frac{d}{d - 2s} \right) \quad \text{and} \quad q \in \left(\frac{d+2s}{d}, \, \frac{d^2 + 2ds + 4s^2}{d^2} \right].
	$$

\noindent	\textbf{Case 1:}  $q \in (\frac{d^2 + 2ds + 4s^2}{d^2}, \, \frac{d}{d-2s})$. In this case, we have
	\begin{equation}\label{q-upperrange}
		\frac{d(q-1)}{2s}    \in \left( \frac{d+2s}{d}, \frac{d}{d-2s}  \right).
	\end{equation}
	
		$\bullet$ First we deal with the non-radial part in \eqref{eq:m-CLR}.  Thanks to  \eqref{q-upperrange}, we have 
	\begin{equation} \label{esti-non-radial-part}
		\int_{\R^d}(P^\perp \widetilde V_N P^\perp)^{\frac{d}{2s}}dx
		\lesssim 
		\int_{\R^d}\rho_{\tilde \gamma_N}^{\frac{d(q-1)}{2s}}dx \lesssim_{d,q,s} 1.
	\end{equation}
	
	$\bullet$
	Next we deal with  the radial part in \eqref{eq:m-CLR}.
	Take $0<s' < s$ close enough to $s$, $s' \neq 1/2$, and $0 <\kappa < \frac{2s'\left(d-1\right)}{\left|1-2s'\right|}$  small enough  such that
	\begin{equation} \label{eq:s-prime}
		\frac{d-2s}{d-2s'} > \frac{d}{d+\kappa}
		\quad
		\text{and }\,
		\frac{d+\kappa}{d-2s'} > \frac{d+2s}{d}\,.
	\end{equation}
	Since $\tr(\cL_s (P\tilde \gamma_NP)) \lesssim_{d,s,q} 1$ (due to \eqref{est:trace(Lgamma)})  and $\cL_s + 1 \gtrsim (-\Delta)^{s'}+1$, it follows that
	\begin{equation} \label{eq:c1-rad-lower}
		\begin{aligned}
			\tr (P \tilde \gamma_N P)
			\gtrsim 
			\tr((\cL_s + 1)(P\tilde \gamma_N P))
			&
			\gtrsim
			\tr(((-\Delta)^{s'}+1)(P\tilde \gamma_N P))
			\\
			&
			\ge
			\|\sqrt{\rho_{P\tilde \gamma_N P}}\|_{H^{s'}}^2
			\gtrsim
			\left(\int_{\R^d}\rho_{P \tilde \gamma_N P}^{\frac{d+\kappa}{d-2s'}}|x|^\kappa dx\right)^{\frac{d-2s'}{d+\kappa}}.
		\end{aligned}
	\end{equation}
	Here the third inequality is due to the Hoffman-Ostenhof  inequality  \cite{HOPRA77} and the last inequality is obtained by applying the fractional Sobolev embedding for radial functions (see \cite[Theorem 6.2]{ND2016}).
	
	Now let us  establish an upper bound for the radial parts  in $B_1$ on the right hand side  of \eqref{eq:m-CLR}. Due to \eqref{q-upperrange}, we can find  $p>1$ such that $  \frac{pd(q-1)}{2s}  \in \left(\frac{d+2s}{d}, \frac{d}{d-2s}   \right)$. Hence, by H\"older's inequality and \eqref{est:rho-Lp}, we  obtain 
	\begin{equation}\label{esti-CLR-|x|-less-1}
		\begin{aligned}
			\int_{B_1} (P \widetilde V_N P)^{\frac{d}{2s}}|\ln|x||^{\frac{d-s}{s}}dx
			\le  \left(  \int_{B_1}   \rho_{P \tilde \gamma_N P}^{\frac{pd(q-1)}{2s}} dx     \right)^{\frac{1}{p}} \left( \int_{B_1}  |\ln|x||^{\frac{p'(d-s)}{s}} dx \right)^{\frac{1}{p'}}
			\lesssim_{d,s,q}  1.
		\end{aligned}
	\end{equation}

	 Next we will establish an upper bound for the radial part in $B_1^c$ on the right hand side  of \eqref{eq:m-CLR}.  Let 
	$\xi \in (0,1)$ such that
	\begin{equation}\label{eq:xi}
		\xi
		<
		\min\left\{
		\frac{d-2s'}{d+\kappa},
		\left(\frac{d+\kappa}{d-2s'} - \frac{d+2s}{d}\right)^{-1}\left[\frac{d(q-1)}{2s}
		- \frac{d+2s}{d}
		\right]
		\right\},
	\end{equation}
	and $\delta >0$ such that 
	\begin{equation} \label{eq:delta}
		\xi\frac{d+\kappa}{d-2s'} + \delta(1-\xi)
		=
		\frac{d(q-1)}{2s}
		\,
		\Longleftrightarrow
		\,
		\delta
		=
		\frac{1}{1-\xi}\left[\frac{d(q-1)}{2s} - \xi\frac{d+\kappa}{d-2s'}\right].
	\end{equation}
	Due to \eqref{eq:s-prime}, \eqref{eq:xi} and \eqref{eq:delta}, we see that 
	\begin{equation}\label{delta-in domain-q}
		\frac{d+2s}{d} < {\delta < \frac{d}{d-2s}}. 
	\end{equation}
	By using \eqref{eq:delta} and H\"older's inequality, we find that 
 \begin{align*}
		&\int_{B_1^c} (P \widetilde V_N P)^{\frac{d}{2s}}|\ln|x||^{\frac{d-s}{s}}dx \\
		&\quad \leq
		\left(\int_{B_1^c}\rho_{P \tilde \gamma_N P}^{\frac{d+\kappa}{d-2s'}}|x|^\kappa dx \right)^{\xi}
		\left(\int_{B_1^c}\rho_{P \tilde \gamma_N P}^{\delta}
		\,
		|\ln|x||^{\frac{d-s}{s\left(1-\xi\right)}}|x|^{-\frac{\kappa\xi}{1-\xi}} dx\right)^{1-\xi}.
	\end{align*} 
	Thanks to \eqref{est:rho-Lp} and the choice of $\delta$ in \eqref{delta-in domain-q}, we get 
	$$\int_{B_1^c}\rho_{P \tilde \gamma_N P}^{\delta}
	\,
	\left|\ln\left|x\right|\right|^{\frac{d-s}{s(1-\xi)}}
	|x|^{-\frac{\kappa\xi}{1-\xi}} dx
	\lesssim_{d,s,q}
	1\,.$$
	Thus, we  obtain 
	\begin{equation}\label{eq:c1-rad-Bc}
		\int_{B_1^c} (P\widetilde V_N P)^{\frac{d}{2s}}|\ln|x||^{\frac{d-s}{s}}dx
		\lesssim_{d,s,q}
		\left(\int_{B_1^c}\rho_{P \tilde \gamma_N P}^{\frac{d+\kappa}{d-2s'}}|x|^\kappa dx \right)^{\xi}.
	\end{equation}
	Combining estimates \eqref{eq:m-CLR}, \eqref{esti-non-radial-part},  \eqref{esti-CLR-|x|-less-1} and \eqref{eq:c1-rad-Bc}, we  derive that 
	\begin{equation} \label{eq:c1-upper}
		\tr(\tilde \gamma_N) \leq \ran(\tilde \gamma_N)
		\lesssim_{d,s,q}
		1
		+
		\left(\int_{B_1^c}(\rho_{P \tilde \gamma_N P})^{\frac{d+\kappa}{d-2s'}}|x|^\kappa dx \right)^{\xi}.
	\end{equation}
	Hence,  we deduce from  \eqref{eq:c1-rad-lower} and \eqref{eq:c1-upper}  that 
	\begin{equation*}
		\left(\int_{\R^d}(\rho_{P \tilde \gamma_N P})^{\frac{d+\kappa}{d-2s'}}|x|^\kappa dx \right)^{\frac{d-2s'}{d+\kappa}}
		\lesssim_{d,s,q}
		1 + \left(\int_{B_1^c}(\rho_{P \tilde \gamma_N P})^{\frac{d+\kappa}{d-2s'}}|x|^\kappa dx \right)^{\xi},
	\end{equation*}
	which, together with the fact that $\xi < \frac{d-2s'}{2(d+\kappa)}$, leads to 
	\begin{equation*}
		\int_{\R^d}(\rho_{P \tilde \gamma_N P})^{\frac{d+\kappa}{d-2s'}}|x|^\kappa dx \lesssim_{d,s,q} 1.
	\end{equation*}
	So,  we derive from   \eqref{eq:c1-upper} that 
	$$\tr(\tilde \gamma_N) \leq \ran(\tilde \gamma_N)\lesssim_{d,q, s} 1.$$ 
	Thus we obtain \eqref{est:trace-ran} in the case
	$q \in \left(  \frac{d^2 + 2sd +4s}{d^2},   \frac{d}{d-2s}  \right) $. 
	
	\medskip 
	\noindent 
	\textbf{Case 2:}  $q \in \left(\frac{d+2s}{d},\frac{d^2 + 2ds + 4s^2}{d^2}\right]$. In this case, we have  $\frac{d(q-1)}{2s} \in \left(1, \frac{d+2s}{d}\right]$. 
	
	$\bullet$ First, we deal with the non-radial part in \eqref{eq:m-CLR}. Let $0 < \alpha_1 < 1$ such that
	\begin{equation*}
		\frac{2s}{\left(q-1\right)d}
		=
		\frac{\alpha_1}{1} + \frac{1-\alpha_1}{\frac{d+2s}{d}},
	\end{equation*}
	then by the interpolation inequality and estimate \eqref{est:rho-Lp}, we have
	\begin{equation*}
		\|P^\perp\rho_{\tilde \gamma_N} P^\perp\|_{L^{\frac{d(q-1)}{2s}}}
		\leq
		\|P^\perp\rho_{\tilde \gamma_N} P^\perp\|_{L^{1}}^{\alpha_1}
		\|P^\perp\rho_{\tilde \gamma_N} P^\perp\|_{L^{\frac{d+2s}{d}}}^{1-\alpha_1}
		\,
		\lesssim_{d,s,q} 
		\left(\int_{\R^d}\rho_{\tilde \gamma_N} dx\right)^{\alpha_1},
	\end{equation*}
	which implies
	\begin{equation} \label{est:nonrad}
	\int_{\R^d} (P^\perp \widetilde V_N P^\perp)^{\frac{d}{2s}}dx \lesssim _{d,s,q} \left(\int_{\R^d}\rho_{\tilde \gamma_N} dx\right)^{\frac{\alpha_1 d(q-1)}{2s}}.	
	\end{equation}
	
	$\bullet$ Next, we will deal with the radial part in \eqref{eq:m-CLR}. To bound the radial part in $B_1$, we can use an argument similar to the one leading to \eqref{esti-CLR-|x|-less-1} with a different $p$ so that we still have  $  \frac{pd(q-1)}{2s}  \in \left(\frac{d+2s}{d}, \frac{d}{d-2s} \right)$, hence  \eqref{esti-CLR-|x|-less-1} still holds true.
	
	To estimate the radial part in $B_1^c$, we employ a similar argument as in Case 1, but with a different choice of  $s'$, $\kappa$, $\xi$ and $\delta$.  More precisely, we choose $0<s'<s$, $s' \neq 1/2$, and $\kappa>0$ small enough (i.e. close enough to $0$) such that $0 < \kappa < \frac{2s'\left(d-1\right)}{\left|1-2s'\right|}$ and
	\begin{equation}\label{eq:c2-s-prime}
		\frac{d-2s'}{d+\kappa} 
		>
		\max
		\left\{
		\frac{d-2s}{d}\,,
		\frac{d}{d+2s}\,,
		\frac{2s}{d\left(q-1\right)}\,,
		{1 - \frac{d}{d+2s}\left[\frac{d(q-1)}{2s} - 1\right] }
		\right\}
		\,.
	\end{equation}
	Note that $s'$ and $\kappa$ exist since 
	$\left(q-1\right)d/2s > 1$.
	Now we choose $\xi$ such that
	\begin{equation} \label{eq:c2-xi}
		\begin{aligned}
			&{\left[\frac{d+2s}{d} - \frac{d(q-1)}{2s}\right]
				\left[\frac{d+2s}{d} - \frac{d+\kappa}{d-2s'}\right]^{-1}}
			<\xi
			\\
			&
			<\min\left\{
			\left[\frac{d}{d-2s}- \frac{d(q-1)}{2s}\right]
			\left[\frac{d}{d-2s} - \frac{d+\kappa}{d-2s'}\right]^{-1}
			\,,{\frac{d-2s'}{d+\kappa}}
			\right\}.
		\end{aligned}
	\end{equation}
	By direct computations, the choice of $s'$ and $\kappa$ in \eqref{eq:c2-s-prime} ensures that $\xi$ is well-defined. Let $\delta$ be as in \eqref{eq:delta}. Due to \eqref{eq:c2-xi}, 
	$$\frac{d+2s}{d} \leq \delta < \frac{d}{d-2s}.$$
	With such a choice of  $\kappa, \xi, \delta$, the upper bound for the radial part in $B_1^c$ in  \eqref{eq:c1-rad-Bc} still holds true.
	
	Now we combine the estimates of radial and non radial parts. 
	From \eqref{eq:m-CLR}, \eqref{esti-CLR-|x|-less-1}, \eqref{eq:c1-rad-Bc}, \eqref{est:nonrad} and \eqref{eq:c1-rad-lower}, we derive two following inequalities
	\begin{align}\label{eq:ineq-sys-a}
			&
			\int_{\R^d}\rho_{\tilde \gamma_N}dx 
			\lesssim_{d,s,q}
			1 + \left(\int_{\R^d}\rho_{\tilde \gamma_N} dx \right)^{\frac{\alpha_1 d(q-1)}{2s}}
			+
			\left(\int_{\R^d}\rho_{P\tilde \gamma_N P}^{\frac{d+\kappa}{d-2s'}}|x|^\kappa dx\right)^{\xi},
			\\ \label{eq:ineq-sys-b}
			&\left(\int_{\R^d}\rho_{P \tilde \gamma_N P}^{\frac{d+\kappa}{d-2s'}}|x|^\kappa dx \right)^{\frac{d-2s'}{d+\kappa}}
			\lesssim_{d,s,q} 1 + 
			\left(\int_{\R^d}\rho_{\tilde \gamma_N} dx \right)^{\frac{\alpha_1 d(q-1)}{2s}}
			+
			\left(\int_{\R^d}\rho_{P\tilde \gamma_N P}^{\frac{d+\kappa}{d-2s'}}|x|^\kappa dx\right)^{\xi}.
		\end{align}
Since $0< \frac{\alpha_1 d(q-1)}{2s}<1$ and $0<\xi < \frac{d-2s'}{d+\kappa}$ (due to  \eqref{eq:c2-xi}), we deduce from \eqref{eq:ineq-sys-a} and \eqref{eq:ineq-sys-b} that 
$$ \int_{\R^d}\rho_{\tilde \gamma_N}dx  \leq N_1,
$$
where $N_1=N_1(d,s,q)$ is independent of $N$. Thus we obtain \eqref{est:trace-ran} in Case 2. The proof of Lemma \ref{lem-stationary-beta_N} is complete. 
\end{proof}
%
	
Now we are ready to give 	
\begin{proof}[Proof of Theorem \ref{thm:optfinite}]
Let $N_1\in \N$ be the constant in Lemma \ref{lem-stationary-beta_N} and $\tilde \gamma_{N_1}$ be the minimizer for the restricted problem \eqref{variation-N-Hardy} with $N=N_1$ obtained in Lemma \ref{lem-exist-finite-trace-op-Hardy}. By a similar argument as in the proof of Theorem \ref{thm:optimizer-LT}, we can show that 
$$ \frac{\|\tilde  \gamma_{N_1}\|_{\dot{\cQ}^s}^\theta}{\| \rho_{\tilde \gamma_{N_1}}\|_{L^q}} = \beta_{N_1} = C_{\rm HLT}.
$$
Therefore, $\tilde \gamma_{N_1}$ is a finite-rank minimizer for \eqref{HLT-inequality-}.

The structure of the minimizer follows from Lemma \ref{lem-existe_optimizer-N-hardy}, while the nonexistence part of Theorem \ref{thm:optfinite} is a consequence of  Lemma \ref{lem-stationary-beta_N}. The proof is complete. 
\end{proof}

\appendix
\section{Concentration compactness lemmas}
First, we  recall Lion's concentration compactness lemma in \cite{L1984}. 
\begin{lemma}[Concentration compactness lemma]\label{concentration-compactness-lemma} Let $\{f_n\}_{n \ge 1}$ be a nonnegative, measurable functions satisfying     $\|f_n\|_{L^1} = 1$ for all $n \ge 1$.  Then, there exists a subsequence $\{f_{n_k}\}_{k \ge 1}$ such that  exactly one of the following statements  holds true.
	\begin{itemize}
		\item[(i)] (\textit{compactness}) There exists a sequence $\{y_k\} \subset \R^d$ such that  $\{f_{n_k}(\cdot + y_k)\}_{k\ge 1}$ is tight, i.e. for all $\epsilon >0$, there exists $R =R(\epsilon) \ge 0$ such that 
		\begin{equation}\label{concentrated-sequence}
			\int_{y_k + B_R} f_{n_k}dx  =1 - \epsilon, \quad \forall k \ge 1.
		\end{equation}

		\item[(ii)] (\textit{vanishing}) There holds
		\begin{equation}\label{vanishing}
		\lim_{R \to \infty} \left( \sup_{y} \int_{y + B_R } f_{n_k} dx \right) = 0.
		\end{equation}
		\item[(iii)] (\textit{dichotomy}) There exists $\alpha \in (0, 1)$ such that for any $\varepsilon > 0$, there exist $k_0 \geq 1$ and $f_k^1, f_k^2 \in L_+^1(\R^d)$ such that  for all $k \geq k_0$,
		\begin{equation}\label{dichotomy-sequence}
			\begin{aligned}
				&\left\|f_{n_k} - \left(f_k^1 + f_k^2\right)\right\|_{L^1} \leq \varepsilon, \\
				&\Big| \int_{\R^d} f_k^1 dx - \alpha \Big| \leq \varepsilon,\quad
				\Big| \int_{\R^d} f_k^2 dx - (1 - \alpha) \Big| \leq \varepsilon,
			\end{aligned}
		\end{equation}
		with
		$$
		{\rm dist}({\rm supp}(f_k^1), {\rm supp}(f_k^2)) \to \infty \quad \text{as } k \to \infty.
		$$
	\end{itemize}
\end{lemma} 

Note that in the above lemma, the compactness is referred to as the tightness of the sequence $\{f_n\}$, which is however not strong enough for our purpose. We will need the following variant of this abstract result, where the compactness in $L^1$-norm is obtained assuming the summability of $f_n$ in an $L^{1+\epsilon}$ space. The idea is somewhat inspired by the ``$pqr$ lemma" in \cite{BFV2013}.

\begin{lemma}\label{concentration-implies-strong-convergence}
	Let   $\{ f_n\}_{n \ge 1}$ be a  sequence of non-negative, measurable functions   in $L^1(\R^d)$ satisfying $f_n \to f$ a.e. in $\R^d$. Assume furthermore that the following properties are valid.
	\begin{itemize}
		\item[(i)] There exist $\epsilon_0>0 $ and $C>0$ such that    
		$$\int_{\R^d} f_n^{1 + \epsilon_0} dx \le C, \quad \forall n \ge 1.$$ 
		\item[(ii)] For all $\epsilon >0$, there exists $R =R(\epsilon) >0$ such that 
		$$  \int_{B_R^c} f_n dx  \le  \epsilon, \quad  \forall n\ge 1
		.$$
	\end{itemize}
	Then
	$$ f_n \to f \text{ strongly  in } L^1(\R^d) \text{ as } n \to \infty.$$
\end{lemma}
\begin{proof}
	By contradiction, we assume that $\{f_n\}$ does not converge to $f$ in $L^1(\R^d)$. Thus there exist $\delta >0$ and a subsequence, still denoted by $\{f_n\}$,  such that 
	$$  \| f_n - f \|_{L^1} \ge \delta,\quad  \forall n \ge 1.$$
	Since $f_n \to f$ a.e. as $n \to \infty$, by Fatou's lemma,  $0 \le f \in L^1(\R^d)$. Moreover, by choosing $\epsilon =  \frac{\delta}{4}$, we can apply  Fatou's lemma again and use the assumptions (i) and (ii) to deduce  that there exists $R=R(\epsilon)$ such that   
	$$ \int_{\R^d} f^{1+\epsilon_0}dx \le C \text{ and }  \int_{B_R^c} f dx \le \frac{\delta}{4}.  $$ 
	Thus, by the triangle inequality, we obtain
	\begin{equation}\label{triangle-inequlaity-f-n-f}
		\int_{B_R} |f_n - f|dx  \ge \frac{\delta}{2}.
	\end{equation}
	Additionally, since $f_n \to f$ a.e. in $\R^d$, we derive from  the  Egorov's theorem  that  for any $\eta >0$,  there exists $\Omega_\eta \subset B_R$ such that  $|\Omega_\eta| \le \eta$ and
	$$ \|f_n - f\|_{L^\infty(B_R \setminus \Omega_\eta)} \to 0 \text{ as } n \to \infty.  $$
	Thus, we infer from \eqref{triangle-inequlaity-f-n-f} that 
	$$\frac{\delta}{2} \le \| f_n - f\|_{L^1(B_R)}    \le |B_R| \|  f_n - f\|_{L^\infty(B_R \setminus \Omega_\eta)} + \|f_n -f\|_{L^{1 +\epsilon_0}}|\Omega_\eta|^{\frac{\epsilon_0}{1+\epsilon_0}} \to 0,$$
	as $n \to \infty$ and $\eta \to 0$, which  leads to a contradiction. The proof is complete.   
\end{proof}

\section{IMS formulas} \label{apendix:IMS}
Let $\gamma$ be a bounded self-adjoint operator on $L^2(\R^d)$ such that $0 < \gamma\le 1$.  Let  $ \chi \in C^\infty_c(\R)$ such that $\chi =1$ on  $B_1$  and  $\chi =0 $ on $B_2^c$. For $R_n \ge 0$,  we define $\chi_n (x) = \chi(\frac{|x|}{R_n})$ and  $\eta_n \ge 0,$ is defined by the relation  $\chi_n^2 + \eta_n^2 =1$ in $\R^d$.

Next, let us recall below the norms $\|\cdot\|_{\dot{\cH}^s}$ and $\|\cdot\|_{\dot{\cQ}^s}$ respectively
\begin{equation*}\label{norm-H-s-dot}
	\|\gamma\|_{\dot{\cH}^s} = \tr\big|(-\Delta)^{\frac{s}{2}}\gamma(-\Delta)^{\frac{s}{2}}\big| \quad \text{and} \quad	\|\gamma\|_{\dot{\cQ}^s} = \tr\big|\sqrt{\cL_s}\gamma \sqrt{\cL_s}\big|.
\end{equation*}

\begin{lemma}\label{lemma-IMS}
Assume $d \geq 1$, $s \in (0,1], s<d/2$. Let $\{\gamma_n\}_{n \ge 1}$  be a sequence of nonnegative, bounded, self-adjoint operators on $L^2(\R^d)$ satisfying $\|\gamma_n\|_{\dot{ \cH}^s} \le C $ and $\|\rho_{\gamma_n} \|_{L^q} \le C$ for some $q \in \left[1 +\frac{2s}{d},\frac{d}{d-2s} \right)$, and  $\{R_n\}_{n \ge 1}$ be a positive sequence such that $R_n \nearrow \infty$.  Then    
	\begin{equation}\label{IMS-Laplace}
		\|\gamma_n\|_{\dot{\cH}^s} = \|\chi_n\gamma_n\chi_n\|_{\dot{\cH}^s} + \|\eta_n\gamma_n\eta_n\|_{\dot{\cH}^s} + o(1)_{n \to \infty}. 
	\end{equation} 
	In particular, if we replace the condition $\|\gamma_n\|_{\dot{ \cH}^s} \le C $ by $\|\gamma_{n}\|_{\dot{\cQ^s}} \le C$,  then
	\begin{equation}\label{IMS-Hardy}
		\|\gamma_n\|_{\dot{\cQ}^s} = \|\chi_n\gamma_n\chi_n\|_{\dot{\cQ}^s} + \|\eta_n\gamma_n\eta_n\|_{\dot{\cQ}^s} + o(1)_{n \to \infty}.
	\end{equation}
\end{lemma}
\begin{proof}
	We first prove \eqref{IMS-Laplace} by considering separately two cases: $s = 1$  and $0<s <1$.
	
	\medskip 
	\noindent 
	\textbf{Case 1:} $s=1$ (and $d \geq 3$).  We will  prove  that 
	\begin{align} \label{IMS-H-s-formu-1}
		\|\gamma_n\|_{\dot{\cH^1}} = 
		\|\chi_n \gamma_n \chi_n\|_{\dot{\cH^1}} +  \|\eta_n \gamma_n \eta_n\|_{\dot{\cH^1}} - \int_{\R^d} (|\nabla \chi_n|^2 + |\nabla \eta_n|^2 ) \rho_{\gamma_n}dx .
	\end{align}
	Since   $(-\Delta)^{\frac12} \gamma_n (-\Delta)^{\frac12}$  is  trace class and $\rho_{\gamma_n} \in L^q(\R^d) $,  $\gamma_n$ has the following spectral decomposition
	$$ \gamma_n = \sum_{j=1}^\infty | u_j^{(n)} \rangle\langle u_{j}^{(n)}|,$$
	where $u_j^{(n)} \in \dot{H}^1(\R^d) \cap L^{2q}(\R^d)$.  It follows that
	\begin{equation}\label{express-gamm-H-1}
		\|\gamma_n\|_{\dot{\cH^1}} = \sum_{j=1}^\infty  \|(-\Delta)^{\frac{1}{2}}u^{(n)}_j\|^2_{L^2}.
	\end{equation}
	For $u \in \dot H^1(\R^d)$, we invoke the  IMS formula (which is due to Michael Loss and appeared in \cite{LiebYau88}) to have
	\begin{equation} \label{IMS-norm-s=1Laplace}
	\begin{aligned} 
		&\|(-\Delta)^{\frac{1}{2}}u_j^{(n)}\|^2_{L^2} \\
		&= \|(-\Delta)^{\frac{1}{2}}(\chi_n u_j^{(n)})\|_{L^2}^2 +  \|(-\Delta)^{\frac{1}{2}}(\eta_n u_j^{(n)})\|_{L^2}^2  - \int_{\R^d} (|\nabla \chi_n|^2 + |\nabla \eta_n|^2 )|u_j^{(n)}|^2 dx.
	\end{aligned}
	\end{equation}
	Since $\rho_{\gamma_n} = \sum_{j=1}^\infty |u_j^{(n)}|^2 $, we infer \eqref{IMS-H-s-formu-1}  from \eqref{express-gamm-H-1} and  \eqref{IMS-norm-s=1Laplace}.  
	
	Next, by H\"older's inequality, we get
	\begin{equation*}
		\left| \int_{\R^d} \left(|\nabla \chi_n|^2 + |\nabla \eta_n|^2 \right) \rho_{\gamma_n} dx \right|  \le   CR_{n}^{\frac{d}{q'}-2}  \|\rho_{\gamma_n}\|_{L^q} \to 0 \text{ as } R_n \to \infty.
	\end{equation*}
	This and \eqref{IMS-H-s-formu-1} imply \eqref{IMS-Laplace}.

	\medskip
	\noindent 
	$\bullet$ \textbf{Case 2:} $0< s < 1$ (and $s<d/2$).  Let $u \in \dot H^{s}(\R^d)$.  By the  IMS formula (see \cite{LiebYau88}, \cite{FLSJAMS08}), 
	\begin{equation} \label{IMS-norm-s small 1-Laplace}
		\|(-\Delta)^{\frac{s}{2}}u\|^2_{L^2} = \|(-\Delta)^{\frac{s}{2}}(\chi_n u)\|_{L^2}^2 +  \|(-\Delta)^{\frac{s}{2}}(\eta_n u)\|_{L^2}^2  - \langle u,\mathscr{H}u\rangle,
	\end{equation} 
	where $\mathscr{H} $ is a bounded operator on $L^2(\R^d)$ whose kernel is determined as follows
	\begin{equation*}
		\mathscr{H}(x,y) = a_{d,s} \frac{\left( \chi_n(x) - \chi_n(y)  \right)^2 +(\eta_n(x) - \eta_n(y))^2 }{|x-y|^{d+2s}}.
	\end{equation*}
	Let us define 
	$$I_{j,k}= \{ z \in \R^d:  jR_n \le |z| \le kR_n  \}.$$
	Then we  can bound the kernel $\mathscr{H}(x,y)$  as  
	$$\mathscr{H}(x,y) \lesssim_{d,s}  H_1(x,y) + H_2(x,y) + H_3(x,y),      $$
	where 
	\begin{equation*}
	\begin{aligned}
		H_1(x,y) &= \frac{1}{R_n^2} \frac{1}{|x-y|^{d+2s-2}} \left(  \mathbbm{1}_{x \in I_{0,1},y \in I_{1,2}} + \mathbbm{1}_{x \in I_{1,2}, y \in I_{0,3}} +\mathbbm{1}_{x \in I_{2,3}, y \in I_{1,2}} \right),\\
		H_2(x,y) &= \frac{1}{(1+|y|)^{d+2s}}  \left(  \mathbbm{1}_{x \in I_{0,1},y \in I_{2,\infty}} + \mathbbm{1}_{x \in I_{1,2}, y \in I_{3,\infty}}\right),\\
		H_3(x,y) &=  \frac{1}{(1 + |x|)^{d+2s}} \left(  \mathbbm{1}_{x \in I_{2,3},y \in I_{0,1}} + \mathbbm{1}_{x \in I_{3,\infty}, y \in I_{0,2}} \right).
	\end{aligned}
	\end{equation*}
	Then, by the  symmetry on $x$ and $y$,  we arrive at
	\begin{equation} \label{est:uHu}
		\left| \langle u, \mathscr{H} u \rangle \right| \lesssim  R_n^{-2}\int_{I_{0,3}} \int_{I_{0,3}} \frac{|u(x)||u(y)|}{|x-y|^{d+2s-2}}   dx dy + \int_{I_{0,3}} \int_{I_{0,3}^c} \frac{|u(x)||u(y)|}{(1 + |y|)^{d+2s}}   dx dy.
	\end{equation}  	
	By using  the Hardy--Litlewood--Sobolev inequality and H\"older's inequality, the first   integral on the right hand side of \eqref{est:uHu} can be estimated by 
	\begin{align}
		\int_{I_{0,3}} \int_{I_{0,3}} \frac{|u(x)||u(y)|}{|x-y|^{d+2s-2}}   dx dy  \lesssim_{d,s} \|u\|_{L^\frac{2d}{d-2s+2}(B_{3R_n})}^2 \lesssim_{d,s} \|u\|_{L^{2}(B_{3R_n})}^2 R_n^{2-2s} .\label{esti-int-doul-x-le-3R-n}
	\end{align} 
	For the second integral on the right hand side of \eqref{est:uHu}, we estimate as follows
	\begin{align}
		\int_{I_{0,3}} \int_{I_{0,3}^c} \frac{|u(x)||u(y)|}{|1 + |y||^{d+2s}}   dx dy  \le \int_{B_{3R_n}} |u(x)|^2dx \int_{B_{3 R_n}^c} \frac{dy}{(1+|y|)^{d+2s}} +  \int_{B_{3 R_n}^c} \frac{|u(y)|^2dy}{(1+|y|)^{d+2s}}.\label{esti-int-doul-x-le-3R-n-y-ge-3R-n} 
	\end{align} 
	By  replacing  $u$ by $u^{(n)}_j$ in \eqref{IMS-norm-s small 1-Laplace} and summing over $j \geq 1$,   we have 
	\begin{align*}
		\|\gamma_n\|_{\dot{\cH}^s} &= \|\chi_n \gamma_n \chi_n\|_{\dot{\cH}^s} +  \|\eta_n \gamma_n \eta_n\|_{\dot{\cH}^s} - \sum_{j=1}^{\infty } \langle u_j^{(n)}, \mathscr{H} u_j^{(n)}\rangle .
	\end{align*}
	Additionally, by using the expression 
	$\rho_{\gamma_n} = \sum_{j=1}^\infty |u^{(n)}_j|^2 $, the estimates \eqref{esti-int-doul-x-le-3R-n}, \eqref{esti-int-doul-x-le-3R-n-y-ge-3R-n} and the assumption $q < \frac{d}{d-2s}$, we  obtain 
	\begin{align*}
		\Big| \sum_{j=1}^{\infty } \langle u_j^{(n)}, \mathscr{H} u_j^{(n)}\rangle  \Big|  &\le C \left(  R^{-2s}_n \int_{B_{3R_n}}  \rho_{\gamma_n}dx  + \int_{B_{3R_n}^c} \frac{\rho_{\gamma_n}dx}{(1 +|x|)^{d+2s}} \right) \\
		&\le C \|\rho_{\gamma_n}\|_{L^q} R_{n}^{\frac{d}{q'} -2s} \to 0 \text{ as } n \to\infty.
	\end{align*}
	Therefore, 
	we derive \eqref{IMS-Laplace} for all  $s \in (0,1]$.
	
	The proof of \eqref{IMS-Hardy} is similar to that of \eqref{IMS-Laplace}, making additionally use of  the identity
	$$
	\frac{1}{|x|^{2s}} = \frac{\chi_n^2}{|x|^{2s}} + \frac{\eta_n^2}{|x|^{2s}}
	$$
	to handle the Hardy potential, hence we omit the proof of \eqref{IMS-Hardy}.
\end{proof}	

The IMS formula can be used to establish the relative compactness of minimizing sequences for the variational problem \eqref{variation-N-Hardy}, which involves the Hardy potential. More precisely, we obtain the following result.

\begin{lemma}\label{lmm:bounded-translation-Ls}
	Let $\left\{\gamma_n\right\}_{n\geq 1}$ be a  minimizing  sequence for  the variational problem \eqref{variation-N-Hardy} with $N$ large enough,  such that
	\begin{equation} \label{asump:gamma_k}
		\left\|\gamma_n\right\|_{\rm op} = 1, \quad \left\|\rho_{\gamma_{n}}\right\|_{L^q} = 1,
		\text{ \rm and }
		\left\|\gamma_{n}\right\|_{\dot{\cQ}^s} \to \beta_N \text{ as } n \to +\infty.
	\end{equation}
	Suppose that  there exists a subsequence  \(\left\{\rho_{\gamma_{k}}\right\}_{k \geq 1}\) and a sequence of points \(\left\{y_{k}\right\}_{k \geq 1} \subset \R^d\) such that
	\[
	\lim_{R \to \infty} \int_{y_k + B_R} \rho_{\gamma_k}^q dx = 1,  \quad \forall k \ge 1.
	\]
	Then the sequence $\{y_k\}_{k \geq 1}$ is bounded.
	
\end{lemma}
\begin{remark}
	This result implies that every minimizing sequence of \eqref{HLT-inequality} is relatively compact.
\end{remark}
\begin{proof}
	Let    $\gamma_\infty$  be  the minimizer for the Lieb--Thirring inequality \eqref{LT-inequality} obtained in Theorem  \ref{thm:optimizer-LT} and recall that $C_{\rm LT}$ and $C_{\rm HLT}$ are the sharp constants in \eqref{LT-inequality} and \eqref{HLT-inequality-} respectively. Then  
	\begin{equation*}
		C_{\rm LT} = \frac{\|\gamma_\infty\|_{op}^{1 -\theta} \| \gamma_\infty\|_{\dot{\cH}^s}^\theta }{ \left\| \rho_{\gamma_\infty} \right\|_{L^q}}
		>
		\frac{\|\gamma_\infty\|_{op}^{1 -\theta} \| \gamma_\infty\|_{\dot{\cQ}^s}^\theta }{ \left\| \rho_{\gamma_\infty} \right\|_{L^q}}
		\geq C_{\rm HLT}.
	\end{equation*}
Note that due to Lemma  \ref{lem_staionary-seq}, $\gamma_\infty$ is also  a minimizer for the variational problem \eqref{variation-N}  with $N$  large enough. Thus, for $N$ large enough, we have 
	\begin{equation} \label{alpha_N>beta_N} \alpha_N  > \beta_N.
	\end{equation}
	
	We suppose by contradiction that there exist a subsequence of $\{y_k\}_{k\geq 1}$, still denoted by the same notation, such that $|y_k| \to +\infty$ as $k \to \infty$. 
	Then, for any $k \geq 1$, put $R_k = \frac{\left|y_k\right|}{4} \to \infty$ and define nonnegative functions $\tilde \chi_k, \tilde \eta_k$ by $\tilde{\chi}_k(y) = \chi(\frac{\left|y- y_k\right|}{R_k})$ and $\tilde{\eta}_k^2 + \tilde{\chi}_k^2 = 1$.
	The decomposition \eqref{IMS-Hardy} gives
	\begin{equation} \label{est:gamma_k}
		\begin{aligned}
			\|\gamma_k\|_{\dot{\cQ}^s}
			&
			=
			\|\tilde{\chi}_k\gamma_k\tilde{\chi}_k\|_{\dot{\cQ}^s}
			+
			\|\tilde{\eta}_k\gamma_k\tilde{\eta}_k\|_{\dot{\cQ}^s}
			+
			o(1)_{k \to \infty}
			\\
			&
			\geq
			\|\tilde{\chi}_k\gamma_k\tilde{\chi}_k\|_{\dot{\cH}^s} 
			-
			\int_{\R^d}
			\frac{\rho_{\gamma_k}(x)\tilde{\chi}_k^2(x)}{|x|^{2s}}dx  + o(1)_{k \to \infty}.
		\end{aligned}
	\end{equation}
	By  H\"older's inequality and the fact that $q'>\frac{d}{2s}$,  we obtain
	\begin{equation*}
		\int_{\R^d}
		\frac{\rho_{\gamma_k}(x)\tilde{\chi}_k^2(x)}{|x|^{2s}}dx \leq
		\left(\int_{\R^d}\rho_{\gamma_k}^qdx\right)^{\frac{1}{q}}
		\left(\int_{B_{R_k/2}^c} |x|^{-2sq'}dx\right)^{\frac{1}{q'}}
		=
		R_k^{\frac{d}{q'}-2s}\left\|\rho_{\gamma_k}\right\|_{L^q} = o(1)_{k \to \infty}.
	\end{equation*}	
	Plugging it back to \eqref{est:gamma_k} and using the assumption \eqref{asump:gamma_k} and the estimate \eqref{variation-N}, we get 
	\begin{equation*}
		\begin{aligned}
			\left\|\gamma_k\right\|_{\dot{\cQ}^s}
			&
			\geq 
			\left\|\tilde{\chi}_k\gamma_k\tilde{\chi}_k\right\|_{\dot{\cH}^s} + o(1)_{k \to \infty}
			\\
			&
			\geq
			\alpha_N \left(\int_{\R^d}\rho_{\gamma_k}^q\tilde{\chi}_k^{2q}dx\right)^{\frac{1}{q}}
			+o(1)_{k \to \infty}
			\to \alpha_N \text{ as } k \to \infty,			\end{aligned}
	\end{equation*}
	which contradicts \eqref{alpha_N>beta_N} since $\left\|\gamma_k\right\|_{\dot{\cQ}^s}  \rightarrow  \beta_N$ as $k\to \infty$.  Thus the sequence $\{y_k\}_k$ is bounded.
\end{proof}

\newcommand{\etalchar}[1]{$^{#1}$}
\providecommand{\bysame}{\leavevmode\hbox to3em{\hrulefill}\thinspace}
\providecommand{\MR}{\relax\ifhmode\unskip\space\fi MR }
\providecommand{\MRhref}[2]{%
  \href{http://www.ams.org/mathscinet-getitem?mr=#1}{#2}
}
\providecommand{\href}[2]{#2}


\newcommand{\etalchar}[1]{$^{#1}$}
\providecommand{\bysame}{\leavevmode\hbox to3em{\hrulefill}\thinspace}
\providecommand{\MR}{\relax\ifhmode\unskip\space\fi MR }
\providecommand{\MRhref}[2]{%
  \href{http://www.ams.org/mathscinet-getitem?mr=#1}{#2}
}
\providecommand{\href}[2]{#2}
\begin{thebibliography}{HOHO77}

\bibitem[BFV13]{BFV2013}
J.~Bellazzini, R.~L. Frank, and N.~Visciglia, \emph{Maximizers for
  {G}agliardo--{N}irenberg inequalities and related non-local problems},
  Mathematische Annalen \textbf{360} (2013), 653--673.

\bibitem[BM24]{BM2024}
K.~Bogdan and K.~Merz, \emph{Ground state representation for the fractional
  {L}aplacian with {H}ardy potential in angular momentum channels}, Journal de
  Mathématiques Pures et Appliquées \textbf{186} (2024), 176--204.

\bibitem[CGW25]{CGWHLT2025}
Bin Chen, Yujin Guo, and Shuang Wu, \emph{Optimizers of the finite--rank
  {Hardy--Lieb--Thirring} inequality for {Hardy--Schr{\"o}dinger} operator},
  arXiv preprint arXiv:2509.17307 (2025), 26.

\bibitem[Cwi77]{Cwikel77}
M.~Cwikel, \emph{Weak type estimates for singular values and the number of
  bound states of {S}chr{\"o}dinger operators}, Annals of Mathematics
  \textbf{106} (1977), no.~1, 93--100.

\bibitem[CX97]{Chemin-Xu97}
J.-Y. Chemin and C.-J. Xu, \emph{Inclusions de {S}obolev en calcul de
  {W}eyl-{H}\"ormander et {C}hamps de vecteurs sous-{E}lliptiques}, Annales
  Scientifiques de l'École Normale Supérieure \textbf{30} (1997), 719--751.

\bibitem[DFL{\etalchar{+}}24]{DFTNT2024}
Giao~Ky Duong, Rupert~L. Frank, Thi Minh~Thao Le, Phan~Th{\`a}nh Nam, and
  Phuoc-Tai Nguyen, \emph{{C}wikel--{L}ieb--{R}ozenblum type inequalities for
  {H}ardy--{S}chr{\"o}dinger operator}, Journal de Mathématiques Pures et
  Appliquées \textbf{190} (2024), 103598, 16 pages.

\bibitem[DLL08]{DLL2008}
J.~Dolbeault, A.~Laptev, and M.~Loss, \emph{{L}ieb--{T}hirring inequalities
  with improved constants}, J. Eur. Math. Soc. \textbf{10} (2008), 1121--1126.

\bibitem[DND16]{ND2016}
Pablo~L. De~N{\'a}poli and Irene Drelichman, \emph{Elementary proofs of
  embedding theorems for potential spaces of radial functions}, Methods of
  Fourier Analysis and Approximation Theory, Applied and Numerical Harmonic
  Analysis, Birkh{\"a}user, Cham, 2016, pp.~115--138.

\bibitem[EF91]{EF1991}
A.~Eden and C.~Foias, \emph{A simple proof of the generalized
  {L}ieb--{T}hirring inequalities in one-space dimension}, J. Math. Anal. Appl.
  \textbf{162} (1991), 250--254.

\bibitem[EF06]{EF2006}
T.~Ekholm and R.~L. Frank, \emph{On {L}ieb--{T}hirring inequalities for
  {S}chr{\"o}dinger operators with virtual level}, Communications in
  Mathematical Physics \textbf{264} (2006), 725--740.

\bibitem[FGL21]{FGL21b}
R.~L. Frank, D.~Gontier, and M.~Lewin, \emph{The nonlinear {S}chr{\"o}dinger
  equation for orthonormal functions ii. application to {L}ieb--{T}hirring
  inequalities}, Communications in Mathematical Physics \textbf{384} (2021),
  1783--1828.

\bibitem[FGL25]{FGL-25}
R.~L. Frank, D.~Gontier, and M.~Lewin, \emph{Optimizers for the finite-rank
  {L}ieb--{T}hirring inequality}, American Journal of Mathematics \textbf{147}
  (2025), no.~2, 503--560.

\bibitem[FHJN21]{FHJN-21}
Rupert~L. Frank, Dirk Hundertmark, Michal Jex, and Phan~Thành Nam, \emph{The
  lieb--thirring inequality revisited}, Journal of the European Mathematical
  Society \textbf{23} (2021), no.~8, 2583--2600.

\bibitem[FLS08]{FLSJAMS08}
R.~L. Frank, E.~H. Lieb, and R.~Seiringer, \emph{{H}ardy--{L}ieb--{T}hirring
  inequalities for fractional {S}chr{\"o}dinger operators}, Journal of the
  American Mathematical Society \textbf{21} (2008), no.~4, 925--950.

\bibitem[FLW22]{FLW22}
R.~Frank, A.~Laptev, and T.~Weidl, \emph{{S}chr{\"o}dinger operators:
  Eigenvalues and {L}ieb--{T}hirring inequalities}, Cambridge Studies in
  Advanced Mathematics, Cambridge University Press, Cambridge, 2022.

\bibitem[Fra09]{F2009}
R.~L. Frank, \emph{A simple proof of {H}ardy--{L}ieb--{T}hirring inequalities},
  Communications in Mathematical Physics \textbf{290} (2009), 789--800.

\bibitem[Fra13]{FCIRM13}
\bysame, \emph{Ground states of semi-linear {PDE}s}, CIRM Luminy, 2013, Summer.

\bibitem[Fra20]{Frank20}
\bysame, \emph{The Lieb-Thirring inequalities: Recent results and open problems}, In: Nine mathematical challenges: an elucidation, A. Kechris, et al. (eds.), 45 - 86, Proceedings of Symposia in Pure Mathematics 104, Amer. Math. Soc., Providence, RI, 2021. 

\bibitem[GLN21]{GLQ-21}
D.~Gontier, M.~Lewin, and F.~Q. Nazar, \emph{The nonlinear {S}chr{\"o}dinger
  equation for orthonormal functions i. existence of ground states}, Archive
  for Rational Mechanics and Analysis \textbf{240} (2021), 1203--1254.

\bibitem[HKY19]{HKY2019}
Y.~Hong, S.~Kwon, and H.~Yoon, \emph{Global existence versus finite time blowup
  dichotomy for the system of nonlinear {S}chr{\"o}dinger equations}, Journal
  de Mathématiques Pures et Appliquées \textbf{125} (2019), 283--320.

\bibitem[HOHO77]{HOPRA77}
M.~Hoffmann-Ostenhof and T.~Hoffmann-Ostenhof, \emph{{S}chr{\"o}dinger
  inequalities and asymptotic behavior of the electron density of atoms and
  molecules}, Physical Review A \textbf{16} (1977), no.~5, 1782--1785.

\bibitem[Lev14]{Lev14}
A.~Levitt, \emph{Best constants in {L}ieb--{T}hirring inequalities: {A}
  numerical investigation}, Journal of Spectral Theory \textbf{4} (2014),
  no.~1, 153--175.

\bibitem[Lie76]{Lieb1976CLR}
E.~H. Lieb, \emph{Bounds on the eigenvalues of the {L}aplace and
  {S}chr{\"o}dinger operators}, Bull. Amer. Math. Soc. \textbf{82} (1976),
  751--754.

\bibitem[Lie81]{Lieb1981}
\bysame, \emph{{T}homas--{F}ermi and related theories of atoms and molecules},
  Rev. Mod. Phys. \textbf{53} (1981), 603--641.

\bibitem[Lie83]{Lieb1983}
E.~H. Lieb, \emph{An $l^p$ bound for the {R}iesz and {B}essel potentials of
  orthonormal functions}, Journal of Functional Analysis \textbf{51} (1983),
  159--165.

\bibitem[Lio84]{L1984}
P.-L. Lions, \emph{The concentration-compactness principle in the {C}alculus of
  {V}ariations. the locally compact case, part 1}, Annales de l'Institut Henri
  Poincar{\'e} C, Analyse Non Lin{\'e}aire \textbf{1} (1984), 109--145.

\bibitem[LS10]{LS10}
E.~H. Lieb and R.~Seiringer, \emph{The stability of matter in {Q}uantum
  {M}echanics}, Cambridge University Press, 2010.

\bibitem[LT75]{LiebThirring75}
E.~H. Lieb and W.~Thirring, \emph{Bound for the kinetic energy of fermions
  which proves the stability of matter}, Physical Review Letters \textbf{35}
  (1975), 687--689.

\bibitem[LT76]{LiebThirring76}
E.~H. Lieb and W.~E. Thirring, \emph{Inequalities for the moments of the
  eigenvalues of the {S}chr{\"o}dinger hamiltonian and their relation to
  {S}obolev inequalities}, Studies in Mathematical Physics (E.~H. Lieb,
  B.~Simon, and A.~Wightman, eds.), Princeton University Press, 1976,
  pp.~269--303.

\bibitem[LY88]{LiebYau88}
E.~H. Lieb and H.-T. Yau, \emph{The stability and instability of relativistic
  matter}, Communications in Mathematical Physics \textbf{118} (1988), no.~2,
  177--213.

\bibitem[Nam19]{Nam-review}
P.~T. Nam, \emph{Direct methods to {L}ieb--{T}hirring kinetic inequalities},
  Proceedings of the Workshop on Density Functionals for Many-Particle Systems
  (Singapore), September 2019.

\bibitem[Rob70]{Rob70}
D.~W. Robinson, \emph{Normal and locally normal states}, Communications in
  Mathematical Physics \textbf{19} (1970), 219--234.

\bibitem[Roz72]{Rozen72CLR}
G.~V. Rozenblum, \emph{Distribution of the discrete spectrum of singular
  differential operators}, Izv. Vysš. Učebn. Zaved. Matematika \textbf{1}
  (1972), no.~84, 75--86.

\bibitem[RS72]{RS1975}
M.~Reed and B.~Simon, \emph{Methods of modern mathematical physics, volume ii:
  {F}ourier {A}nalysis, {S}elf-{A}djointness}, Academic Press, 1972.

\bibitem[RS78]{RS1978}
\bysame, \emph{Methods of modern mathematical physics, volume iv: Analysis of
  operators}, Academic Press, 1978.

\bibitem[Rum10]{Rumin2010}
M.~Rumin, \emph{Spectral density and {S}obolev inequalities for pure and mixed
  states}, Geom. Funct. Anal. \textbf{20} (2010), 817--844.

\bibitem[Rum11]{Rumin2011}
\bysame, \emph{Balanced distribution-energy inequalities and related entropy
  bounds}, Duke Math. J. \textbf{160} (2011), 567--597.

\bibitem[Sim79]{BS1979}
B.~Simon, \emph{Trace ideals and their applications}, London Mathematical
  Society Lecture Note Series, vol.~35, Cambridge University Press, Cambridge,
  1979.

\bibitem[SN18]{SNJFA18}
M.~Squassina and H.-M. Nguyen, \emph{Fractional
  {C}affarelli--{K}ohn--{N}irenberg inequalities}, Journal of Functional
  Analysis \textbf{274} (2018), no.~9, 2661--2672.

\end{thebibliography}
\end{document}